\documentclass{article}
\usepackage{graphicx} 
\usepackage{hyperref,verbatim}
\usepackage[normalem]{ulem}
\usepackage[dvipsnames]{xcolor}
\usepackage{latexsym,array,delarray,amsthm, amsmath,amssymb,epsfig,tabu}
\usepackage{bm}
\usepackage{caption}
\usepackage{diagbox}
\usepackage{caption}
\usepackage{subcaption}
\usepackage[ruled,linesnumbered]{algorithm2e}
\usepackage{dsfont}
\usepackage{tikz-cd,tikz}
\usepackage{natbib}

\theoremstyle{plain}
\newtheorem{thm}{Theorem}[section]
\newtheorem{lemma}[thm]{Lemma}
\newtheorem{prop}[thm]{Proposition}

\newtheorem*{thm*}{Theorem}
\newtheorem*{lemma*}{Lemma}
\newtheorem*{prop*}{Proposition}
\newtheorem*{cor*}{Corollary}
\newtheorem*{conj*}{Conjecture}

\theoremstyle{definition}
\newtheorem{defn}[thm]{Definition}
\newtheorem{ex}[thm]{Example}

\newtheorem{rmk}[thm]{Remark}

\theoremstyle{remark}

\newcommand{\R}{\mathbb{R}}
\newcommand{\N}{\mathbb{N}}
\newcommand{\Z}{\mathbb{Z}}

\renewcommand\bar\overline

\DeclareMathOperator{\Ex}{\mathbb{E}}
\DeclareMathOperator{\chibc}{\chi^2_{BC}}

\newcommand{\prob}{\mathbb{P}}
\newcommand{\Fcal}{\mathcal{F}}
\newcommand{\hth}{\hat{\theta}}
\newcommand{\hz}{\hat{z}}

\newcommand{\tB}{\tilde{B}}
\newcommand{\tG}{\tilde{G}}
\newcommand{\tm}{\tilde{m}}
\newcommand{\tn}{\tilde{n}}
\newcommand{\tth}{\tilde{\theta}}

\newcommand{\tz}{\tilde{z}}
\newcommand{\ttee}{\tilde{t}}

\newcommand{\PRVSBM}{\text{LSBM}}

\newcommand{\minus}{\backslash}

\newcommand{\indep}{\perp \!\!\! \perp}

\newcommand{\ep}{\epsilon}

\newcommand{\st}{\text{s.t.}}

\title{
Non-asymptotic goodness-of-fit tests and model selection in  valued stochastic blockmodels} 
\author{F\'elix Almendra-Hern\'andez, Miles Bakenhus,\\ Vishesh Karwa, Mitsunori Ogawa, and Sonja Petrovi\'c\thanks{Authors listed in alphabetical order.}\thanks{
F\'elix Almendra-Hern\'andez, UC Davis, falmendrahernandez@ucdavis.edu\\
Miles Bakenhus, Illinois Institute of Technology, mbakenhus@hawk.illinoistech.edu\\
Vishesh Karwa, Temple University, vishesh@temple.edu\\
Mitsunori Ogawa, Tokyo Medical University, ogawa.mitsunori.6j@tokyo-med.ac.jp\\
Sonja Petrovi\'c, Illinois Institute of Technology, sonja.petrovic@illinoistech.edu 
}}

\begin{document}

\maketitle

\begin{abstract} 
A valued stochastic blockmodel (SBM) is a general way to view networked data in which nodes are grouped into blocks and links between them are measured by counts or labels.  This family allows for varying dyad sampling schemes, thereby including the classical, Poisson, and labeled SBMs, as well as those in which some edge observations are censored.  This paper addresses the question of testing goodness of fit of such non-Bernoulli SBMs, focusing in particular on finite-sample tests. We derive explicit Markov bases moves necessary to generate samples from reference distributions and define goodness-of-fit statistics for determining model fit, comparable to those in the literature for related model families.  
For the labeled SBM, which includes in particular the censored-edge model, we study the asymptotic behavior of said statistics. One of the main purposes of testing goodness-of-fit of an SBM is to determine whether block membership of the nodes influences network formation. Power and Type 1 error rates  are  verified on simulated data. Additionally, we discuss the use of asymptotic results in selecting the number of blocks under the latent-block modeling assumption. The method derived for Poisson SBM is applied to ecological networks of host-parasite interactions. Our data analysis conclusions differ in selecting the number of blocks for the species from previous results in the literature. 
\end{abstract} 


\section{Introduction}
Analysis of network data provides insights into the structure and dynamics of complex systems, generating interest in the development of statistical models for networks. The foundation of probabilistic modeling of network data lies, of course, in the classical random graph model, the Erdös-Rényi-Gilbert model \cite{ErdosRenyiRandomGraphs, GilbertRandomGraphs}.  The community or block membership of actors in a network can play a significant role in the way the relationships between nodes in a network are formed. Network models that capture the block effect are built on the basic stochastic blockmodel, popularly known as the SBM.  Originally proposed in the social sciences by \cite{Fienberg1981CategoricalDA}, the basic SBM postulates that the block membership of nodes is {known} and is the main effect for network edge formation. This model has been extended to latent blocks \cite{Holland83SBMs} (see also \cite{nowicki2001estimation}) and is one of the more popular community structure modeling approaches in practice. Over time, the SBM has been extended to accommodate various complexities, including variable degree distributions, mixed block membership, and dynamic networks. These extensions have established SBMs as a central tool in the analysis of network data within statistics, computer science, and machine learning. The comprehensive review by \cite{Goldenberg2010NetworkModels} and the book \cite{Kolaczyk2017} underscore the pivotal role of SBMs in modern research. 

While \cite{AndersonWasserman} build stochastic blockmodels on dyadic interaction data recorded at multiple levels --- with multiple sociomatrices recording different types of interactions --- much of the classical modeling framework for network data uses simple graphs with dyads Bernoulli random variables. In practice, when data consist of interaction counts, such as in neuronal networks (see, e.g., analyses in \cite{GoF-loglinearERGMs}), the data are thresholded to obtain a simple graph. Examples of creating simple graphs from multiple edge networked data permeate the literature; see, for example, the excellent contribution \cite{JiJin-AOAS}, followed by comment articles, including \cite{KarwaAOAS} which reveals a different view of the data when higher-order structures are considered in the model, and \cite{JiJin-comment1} which discusses the role of embeddings and combining information from citation and coauthorship networks.
Not strictly within the SBM literature, contemporary work on modeling network data generalizes the simple graph setting by allowing multiple connections between nodes, possibly of different types. For example, \cite{Krivitsky2012valued} extends the exponential family random graph modeling (ERGM) framework to valued networks whose relationships are unbounded counts, explores various estimation, modeling, degeneracy, and computational issues that arise from such a generalization, and proposes ways to model common network features for count data.  \cite{AleSteveSonjaAOSbeta}  define the ``generalized" $\beta$-model that allows a (bounded) count observation on each dyad of the network.  \cite{SnijdersMultilevelLongitudinal} further discuss modeling of longitudinal data for multi-level networks. 

\emph{Of particular interest to us are instances where the link between two nodes in a network is measured by counts or labels.} Examples of such networks are commonly found in the ecological and social sciences. This has led to adaptations of the SBM that take into account all available information, rather than simply reducing it to the presence or absence of an interaction. These variants of the SBM are collectively referred to as \emph{valued SBMs}. We define the valued SBM formally in Section~\ref{section:block-models-valued} and note it includes the classical, Poisson, and censored blockmodels. 

\emph{This paper addresses the question of model goodness-of-fit for non-Bernoulli SBMs.}   The question of goodness of fit of network models has received a surge of attention  recently.  We single out only the contributions in this area that are relevant to our work; namely, spectral goodness-of-fit tests developed in \cite{Lei2016GoFtestSBM};  many methods for assessing model fit in ERGMs by comparing reference distributions of network statistics \cite{Hunter2008GoFtestSocialNetworkModel}, see also \cite{RSiena}; finite-sample tests for Bernoulli SBMs \cite{GoFSBMvariants2023}; an improved test for network models \cite{Hu-SBM-GoF} directly addressing the computational expense and lack of theoretical guarantees for the null distribution and asymptotic power from the previous citation; and methods for testing inhomogeneous random graph models \cite{Gesine-inhomogeneousRandomGraph-SteinDiscrepancy}. 
The problem of model selection is another crucial aspect of network analysis. For SBMs, this involves, in particular, the step of determining the number of communities in a network, assuming that it follows a blockmodel. Although model selection and goodness-of-fit testing are related, the latter is a more general problem that can also aid in model selection when applied sequentially. In addition, goodness-of-fit tests provide a way to measure the adequacy of the model, providing valuable insight into how well the model captures the underlying structure of the network itself. 
As we worked on finalizing this manuscript, \cite{JJinNetworkGof2025} provided a very timely review of various block-modeling frameworks and the importance of determining the goodness of fit. In particular, they tackle the following question: ``Out of many existing models, which achieves a better balance between practical feasibility/interpretability and
mathematical tractability?" To this end, they propose a new metric for measuring goodness of fit of degree-corrected mixed-membership blockmodels, as they are the most general in the framework. They work with simple undirected graphs, although they also acknowledge insights to dynamic and multi-layer networks as well.  

\emph{We focus on finite-sample tests.} 
The case for finite-sample tests of goodness of fit has been made in the literature, particularly in the case of small-sample data. \cite{TheFienbergAdvantage} offers a discussion of this problem for cross-classified count and relational data, including valued networks. On the topic of theoretical and effective sample size in networks, the reader is likely familiar with \cite{KrivitskyKolaczyk15}. Another excellent discussion about why large-sample criteria for model selection, such as the Bayesian Information Criterion (BIC) and its extensions, fail to properly specify the penalty for various network models whose parameter vectors' dimension increases appeared recently in \cite{Schw2024betaGoF}, who also provides a nonasymptotic GoF test criterion for the $\beta$-model. 

\emph{Our contributions.}
Given their generality, SBMs are amenable to two types of modeling assumptions; namely, that the block assignment for each node can either be fixed or latent. 
In Section~\ref{sec:GoF tests main},  we assume that the block membership is fixed, regardless of whether it is known or not.
When the block membership is known (as in Section~\ref{sec:conditional test}), the goodness-of-fit test is the classical conditional exact test.  We derive Markov bases for Poisson and labeled SBMs, allowing us to present a general MCMC algorithm.
It is well known that when explicit Markov bases are available, the resulting dynamic sampling algorithm provides a valid sample from the conditional distribution for testing. 
In the scenario where block membership is unknown, we will consider both frequentist (in Section~\ref{sec:plug-in z-hat test}) and Bayesian (in Section~\ref{sec:bayesian test}) approaches to  assessing goodness-of-fit of the model.
 
Implementing the test in practice necessitates the development of a nondegenerate goodness-of-fit statistic; this is the content of Section~\ref{sec:GoF statistic}. 
The goodness-of-fit statistic depends on the block membership of the nodes, denoted by $z$ in the following sections, and the model parameters, $\theta$. When $z$ is known,  one can use the MLE $\hat \theta_{mle}$ as a consistent estimator of $\theta$. When $z$ is estimated, we show in Section \ref{section:consistency of mle} that the plug-in estimator $\hat \theta(\hat z)$ is a consistent estimator of $\theta$, as long as $\hat z$ is weakly consistent.  

Section~\ref{section: merged blocks asymptotics} discusses what happens when the number of blocks $k$ is unknown and in particular the implications of  goodness-of-fit tests for model selection. Namely, while we cannot expect to reject overspecified models, we provide theory that supports the selection of $k$ as the minimal $k$ for which the goodness-of-fit tests we develop do not reject the model. The asymptotics of the goodness-of-fit statistics are derived for the labeled SBM. Notably, this model variant includes the censored stochastic blockmodel introduced in \cite{Abbe2013ConditionalRandomFields}; see also \cite{SouvikCSBM2022}. 

Finally, our testing method is demonstrated on simulated data in Section~\ref{sec:type I and power} and model selection for two host-parasite species networks in Section~\ref{sec:selecting k in species networks}. These simulations focus on Poisson SBM for the purpose of comparison.  
We close with a discussion of future work in Section~\ref{sec:future}. 

\section{Blockmodels for valued networks}\label{section:block-models-valued}

The early analyses using SBMs model simple graphs, undirected links between distinct nodes either present or absent with some probability that depends on the nodes' block membership.  
In this section, we propose a general version of blockmodels for \emph{valued networks} and relate them to some well-known blockmodel variants from the existing literature.

\begin{defn}[Valued SBM]\label{def:valued SBM}
Let $n, k\in \Z_{+}$ represent the number of nodes and the number of blocks, respectively. We  consider symmetric, loopless random valued networks $G=(G_{uv})_{1\leq u<v\leq n}$ on the node set $[n]:=\{1, \ldots, n\}$. For simplicity, for a \emph{dyad} or pair of nodes $\{u,v\}$, $G_{uv}$ will represent the entry in $G$ associated with the sorted dyad.
The block assignment of the $n$ nodes will be denoted by $Z=(Z_1, \ldots, Z_n)\in [k]^n$. 

The \emph{Valued SBM} postulates that 
\[
G_{uv} \indep G_{u'v'}\mid Z \; \text{ for any two pairs of {dyads} }\{u,v\} \text{ and } \{u',v'\}.
\]
We define the $i$-th block of the block assignment as $B_i=\{u\in [n]: Z_u=i\}$. Furthermore, we assume the existence of a vector of parameters $\theta=(\theta_{ij})_{1\leq i \leq j \leq k}$ such that
\[
G_{uv}\mid Z=z  \sim f_{uv}(\cdot\;; z, \theta). 
\] 
For each $1\leq u < v \leq n$, $f_{uv}(\cdot\;; z, \theta)$ is a known probability distribution determined by $z$ and $\theta$. We choose the form of this probability distribution depending on the type of network data we are dealing with. In this work, we will assume that for any realization $g$ of $G$, the conditional probability of $G$ given $Z=z$ takes an exponential family form (see also exponential random graph models defined in \cite{Robins2007ERGMs}): 
\begin{equation}\label{eqn:exponential_family}
\mathbb P_\theta(G=g\mid Z=z)=f(g;z, \theta) := \prod_{1\leq u<v \leq n} f_{uv}(g_{uv}; z, \theta) = \prod_{1\leq u<v \leq n} \frac{h(g_{uv})\exp\langle T_z(g_{uv}), \theta_{z_u z_v}\rangle}{\psi(\theta_{z_u z_v})}, 
\end{equation}
where $h(g_{uv})$ is the base measure, $\psi(\theta_{z_u z_v})$ is a normalizing constant, $T_z(g_{uv})$ is a sufficient statistic and $\theta_{ij} = \theta_{z_u z_v}$ is the parameter corresponding to the dyad random variable $G_{uv}$, $u \in B_i$ and $v \in B_j$. 
\end{defn}
In other words, the valued SBM postulates that conditional on $Z=z$, (a) the dyad random variables $g_{uv}$ are independent, and (b) the distribution of each dyad random variable, given by $f_{uv}(g_{uv}; z, \theta)$, belongs to an exponential family model with (possibly vector valued)  natural parameter $\theta_{ij}$. Hence, conditional on $z$, the distribution $f(g;z,\theta)$  of $g$ can be factorized as the product of $f_{uv}(g_{uv};z,\theta) $. The choice of $f_{uv}$ will determine the state space for $G$ that will be denoted in general by $\mathbb G$.
When $z$ is known, Equation \eqref{eqn:exponential_family} can be further simplified  (but the reader should note that this simplification assumes a one parameter exponential family for the dyad $G_{uv}$; for a vector-valued family, additional notation is needed). Define $$T_{z,ij}(g) = \sum_{u \in B_i v \in B_j} T_z(g_{uv}), \quad h_{z,ij}(g) = \prod_{u \in B_i v \in B_j} h(g_{uv}), \mbox{ and } \psi_{z,ij}(\theta_{ij}) = \prod_{u \in B_i v \in B_j} \psi(\theta_{uv}) $$ then, we have:
\begin{equation}\label{eqn:exponential_family_known_z}
\mathbb P_\theta(G=g\mid Z=z)= \prod_{i,j}\prod_{u \in B_i v \in B_j} \frac{h(g_{uv})\exp\langle T_z(g_{uv}), \theta_{uv}\rangle}{\psi(\theta_{uv})} = \exp\left(\sum_{ij}\theta_{ij} T_{z,ij}(g)\right) \prod_{ij}\frac{h_{z,ij}(g)}{\psi_{z,ij}(\theta_{ij})}. 
\end{equation}
This is an exponential family model with vector of sufficient statistics $T_z(g) = \{T_{z,ij}(g)\}$, vector of parameters $\theta = \{\theta_{ij} \}$, base measure $h_z(g) = \prod_{ij} h_{z,ij}(g)$ and normalizing constant $\psi_z(\theta) = \prod_{ij} \psi_{z,ij}(\theta_{ij})$.
When $z$ is unknown, the model is no longer an exponential family model. In such a case, we treat $z$ as a fixed but unknown parameter. However, we are abusing the notation and use $\mathbb P_{\theta}(G=g\mid Z)$ instead of $\mathbb P_{\theta,z}(G=g)$ to denote the unknown $z$ case. This is because in the unknown $z$ case, we construct two tests: one that uses a plug-in estimate of $z$ and another that takes into account the uncertainty in the estimate of $z$ by assuming $Z$ to be a random variable and using a posterior distribution of $Z$. Both tests rely on the conditional distribution $P_{\theta}(G=g\mid Z=z)$ being an exponential family form.

The generality of the valued SBM allows us to consider several well-known models as special cases, by specifying a dyad sampling scheme. Three examples of particular interest to us follow.

\subsection{Classical SBM}\label{subsection:classicSBM}

Specifying $G_{uv}\mid Z=z \sim \text{Bernoulli}(\theta_{z_uz_v})$ retrieves the classical SBM from \cite{Holland83SBMs} for which every $G_{uv}$ determines the existence or absence of an interaction between nodes $u$ and $v$. In turn, this means that $\mathbb G =\{0,1\}^{\binom{n}{2}}$ and $\theta_{ij}\in (0,1)$ represents the probability of having an interaction between nodes in blocks $i,j\in[k]$. The base measure for this setting is $h\equiv 1$ and given $Z=z$, the vector of sufficient statistics is $T_{z}(g)=(T_{z, ij}(g): 1\leq i\leq j \leq k)$, where
    \begin{equation} \label{eq:suff_stat_bernoulli}
T_{z, ij}(g) = \begin{cases} \sum\limits_{u \in B_i, v\in B_j}g_{uv}, & \text{ if } i\neq j \\ \frac{1}{2} \sum\limits_{u\neq v\in B_i} g_{uv}, & \text{ if } i=j.\end{cases}
\end{equation}
In other words, $T_{z, ij}(g)$ is the total number of interactions between blocks $B_i$ and $B_j$.

\subsection{Poisson SBM}\label{subsection:poissonSBM}

Let $G_{uv}\mid Z=z\sim \text{Poisson}(\theta_{z_uz_v})$. This model is known as the Poisson SBM in the literature (see \cite{mariadassou2010valued}, \cite{Signorelli2018ItalianParliament}).  Here, we interpret  the values $G_{uv}$ as the number of interactions between nodes $u$ and $v$, yielding $\mathbb G = \mathbb N^{\binom{n}{2}}$. For each $i\leq j$, $\theta_{ij}\in \R_+$ represents the average number of interactions between nodes in blocks $i,j$ and for each realization $g\in \mathbb G$, $h(g)=\prod_{1\leq u<v \leq n}\frac{1}{g_{uv}!}$ and $T_{z}(g)$ is defined as in \eqref{eq:suff_stat_bernoulli}.  

\subsection{Labeled SBM}\label{subsection:labeledSBM}

Let $G_{uv}\mid Z=z \sim \text{Multinomial}\left(N, (\theta^{(\ell)}_{z_uz_v})_{\ell=1}^L\right)$. This model is known as the Labeled SBM in the literature (see \cite{Yun2016LabeledSBM}, \cite{heimlicher2012LabelledSBM}). In this case, we consider $L$ different interaction (or edge) types between nodes. This means that for each dyad $\{u,v\}$, $G_{uv}$ is an $L$-dimensional vector $G_{uv}=(G_{uv}^{(\ell)})_{\ell=1}^L$ with $G_{uv}^{(\ell)}$ being the number of $\ell$-type interactions (or edges) between nodes $u$ and $v$. Here, $N$ is the total number of existing interactions between any pair of nodes and $\theta_{ij}^{(\ell)}\in (0,1)$ represents the probability that an interaction between nodes in blocks $i$ and $j$ is of type $\ell$. Hence, we have the constraint 
    \begin{equation}\label{eqn:multinomial_params_contraint}
    \sum_{\ell=1}^L \theta_{ij}^{(\ell)}=1 \text{ for every } 1\leq i\leq j\leq k.
    \end{equation}

For this model, the valued graphs state space is given by $\mathbb G = \{g\in \N^{L\binom{n}{2}}:\sum_{\ell=1}^L g_{uv}^{(\ell)}=N\}$, with base measure given by $h(g)=\prod_{u<v}\frac{N_{uv}!}{g_{uv}^{(1)}! \cdots g_{uv}^{(\ell)}!}$, and vector of sufficient statistics given by $T_{z}(g)=(T_{z,ij}^{(\ell)}(g):1\leq i\leq j\leq k, \ell\in [L])$ where
\begin{equation} \label{eq:suff_stat_multinomial}
T_{z, ij}^{(\ell)}(g) = \begin{cases} \sum\limits_{u \in B_i, v\in B_j}g^{(\ell)}_{uv}, & \text{ if } i\neq j \\ \frac{1}{2} \sum\limits_{u\neq v\in B_i} g^{(\ell)}_{uv}, & \text{ if } i=j.\end{cases}
\end{equation} 

\begin{rmk} The classical SBM is recovered from this model as a special case, by setting $\ell=2, N=1$. Since $G_{uv}^{(2)}=1-G_{uv}^{(1)}$ for every dyad $\{u,v\}$, we have that $G_{uv}^{(1)}\mid Z=z \sim \text{Bernoulli}(\theta_{z_uz_v}^{(1)})$. Then, instead of considering the full vector $(G_{uv}^{(1)}, G_{uv}^{(2)})$, one can simply consider the random variable $G_{uv}:=G_{uv}^{(1)}$ for each dyad $\{u,v\}$.
\end{rmk}

\begin{ex}
    [Modeling censored network data] The Censored Stochastic Blockmodel considered in \cite{Abbe2014CensoredSBM}, \cite{SouvikCSBM2022} and introduced in \cite{Abbe2013ConditionalRandomFields} in a different context, considers simple graphs on $n$ vertices with a latent vertex-block assignment $Z$. Two vertices $u\neq v$ are connected by an edge with probability $q_{11}$ if they both belong to community $1$, probability $q_{22}$ if they both belong to community $2$ and $q_{12}=q_{21}$ if they belong to different communities. Finally, each status is revealed independently with probability $\alpha$. The output is a graph with dyad states given by \emph{present}, \emph{absent} or \emph{censored}, which we represent with the values $1, 2$ and $3$, respectively.

This means that if $G=(G_{uv}:1\leq u<v \leq n)$ is a random valued network, under the Censored SBM model we have
\begin{equation}
\mathbb P(G_{uv}=\bm e_\ell\mid Z=z)=
\begin{cases}
\alpha q_{z_uz_v} & \text{if } \ell=1 \;(\text{present}),\\  
\alpha (1-q_{z_uz_v}) & \text{if } \ell=2\;(\text{absent}),\\  
1 - \alpha & \text{if } \ell=3 \;(\text{censored}). 
\end{cases} 
\end{equation}
Here, $\bm e_{\ell}$ represents the $\ell$-th $3$-dimensional unit vector. We are able to recover the Censored SBM from the Labeled SBM by setting $k=3, N=1$ and $\theta_{ij}^{(1)}=\alpha q_{ij}, \theta_{ij}^{(2)}=\alpha (1-q_{ij}), \theta_{ij}^{(3)}=1-\alpha$. 
\end{ex}

\section{Testing goodness-of-fit of  valued SBMs}\label{sec:GoF tests main}

Given a valued network $g\in \mathbb G$ and a  number of blocks $k\in \Z_+$, we would like to know if $g$ can be modeled by a valued SBM  with a  block assignment that partitions the nodes of $g$ into $k$ blocks. 
Since there are two possible cases depending on whether a reasonable block assignment is known or not, we build two versions of the test: one for known and one for unknown block assignment.

In this section, we specify the goodness-of-fit testing hypotheses and assumptions, and discuss how each of the tests is carried out in practice. Each model variant requires the use of a valid goodness-of-fit statistic (or discrepancy measure as discussed in \cite{Meng94}) to evaluate model fit; this discussion is deferred to Section~\ref{sec:GoF statistic}.

As for any exponential family model, we can condition on sufficient statistics to define a classical exact conditional test for goodness-of-fit. This is the context of \cite{GoFSBMvariants2023}, which further relies on the fact that the Markov bases machinery, discussed in this section, can be used to effectively  sample from the conditional distributions. We follow a similar argument for labeled SBMs, first deriving the tools necessary for testing the model under a fixed block assignment, then extending that approach to the unknown block assignment case.

\subsection{Conditional goodness-of-fit test under fixed block assignment}
\label{sec:conditional test}

When the random valued network $G$ is generated using a fixed block assignment $z$, we consider the following goodness-of-fit test. We test the null hypothesis that $G$ arises from the SBM, 
\[
H_0:G\sim \mathbb P_\theta(G\mid Z=z),
\] 
with $\theta\in\Theta$ and a fixed $z$, 
against the general alternative. Since $P_\theta(G\mid Z=z)$ belongs to the exponential family, a natural conditional test for $H_0$ is to condition on its sufficient statistic to remove the dependence on the unknown model parameters.

\begin{defn} Let $z$ be a fixed block assignment and $T_z(\cdot)$ be the sufficient statistic of the exponential family $f(\cdot \;; z, \theta)$ as in \eqref{eqn:exponential_family}. We define the following subset of the sample space:

\begin{equation}
\mathcal F_{z,t}:=\{g\in \mathbb G:T_z(g)=t\}, \label{eq:fiber}
\end{equation}
called the \emph{fiber of $(z,t)$} under the valued SBM. This set is the support of the conditional distribution given the block assignment $z$ and the sufficient statistics $t$.
\end{defn}

When the random valued network $G$ is generated using a fixed block assignment $z$, any goodness-of-fit statistic $\text{GoF}_{z}(g)$ that is a function of the valued network $g$ and the block assignment $z$, such that large values of $\text{GoF}_{z}(g)$ imply departures from the model, leads to an exact conditional $p$-value
\begin{equation}\label{eqn:frequentist-p-value}
p(z, g)=\mathbb P(\text{GoF}_{z}(G)\geq \text{GoF}_{z}(g)\;|\; T_{z}(g)).
\end{equation}

Note that, when considering the hypothesis $H_0$ above,  knowing whether $z$ is the true block assignment or not does not affect the validity of the corresponding exact conditional test. (The reader should refer to Section~\ref{sec:discussion} for a discussion of the meaning and interpretation of a `true' block assignment.) 
Even if the true $z^*$ is different from $z$ specified in the hypothesis $H_0$, the specified $z$ can be used as the true block assignment under $H_0$, which gives the validity of the test with respect to the type I error. 

\smallskip 
For completeness, we provide the conditional distribution on the fiber defined in Equation~\eqref{eq:fiber}. 
\begin{lemma}[Cf.\ Lemma 1 in \cite{newDirections2024}]
    Let $g\in \mathbb G$ denote a graph,  and let $\mathbb P_\theta(G=g)$ be defined as in~\eqref{eqn:exponential_family}, defining an exponential family model on $g$ where $h(g)$ is the base measure, $\theta$ is a vector of parameters, $T_z(g)$ is the vector of sufficient statistics. Then 
\begin{align}
    \mathbb P\left(G=g\mid T_z(g)=t\right) = \frac{h(g) }{\sum_{g'\in\mathcal F_{z,t}}h(g')},  
\end{align}
where $\mathcal F_{z,t}$, defined in \eqref{eq:fiber}, is the set of graphs $g\in\mathbb G$ whose sufficient statistics are equal to $T_z(g)$. 

Under the Poisson and multinomial sampling schemes on the dyads, this conditional distribution on the fiber is hypergeometric. Under geometric or Bernoulli sampling scheme on the dyads, the conditional distribution on the fiber is the uniform distribution. 
\end{lemma} 

Notice that $\theta$ is not included in the notation of the conditional distribution $\mathbb P_\theta(G\;|\;T_z(g))$ since for a fixed sufficient statistic value $t$
\[
\mathbb P_\theta(G=g\mid T_{z}(G)=t)=\frac{h(g)}{\sum_{g'\in \mathcal{F}_{z,t}}h(g')} \text{ if }g\in \mathcal F_{z,t}, \;\text{ and } 0 \text{ otherwise}.
\] 
Since enumerating the fiber $\mathcal F_t$ is computationally intractable, one usual approach is to sample from it using a Markov chain Monte Carlo algorithm in order to approximate the $p$-value stated in \eqref{eqn:frequentist-p-value}. As is well known by now, the Fundamental Theorem of Markov Bases \cite{DS98} states that when $T_{z}(g)$ is linear on $g$ (such as in the case of the models listed in Section \ref{section:block-models-valued}), there \emph{always} exists a finite set of steps, or moves, that one can use to sample from the conditional distribution on the fiber. This set is called a \emph{Markov basis}, and is guaranteed to connect all fibers of a given log-linear exponential family model. 
To define it, note that $T_z(g)$ being a linear operation means there exists a \emph{configuration matrix} $A_{T_z}$ such that $T_z(g)=A_{T_z}\cdot g$, where $g$ has been flattened to a vector; this matrix is sometimes called a design matrix in the study of log-linear models. 

\begin{defn}[Markov basis]\label{defn:Markov_basis}  Let $z$ be a fixed block assignment and $T_z(\cdot)$ be the sufficient statistic of the exponential family $f(\cdot \;; z, \theta)$. Let $\mathcal B$ be any set of vectors in $\ker_{\Z}(A_{T_z})$.  The set $\mathcal B$ is said to connect the fiber $\mathcal F_{z,t}$ if given any two graphs $f, g \in \mathcal F_{z,t}$, there exist moves $b_1, \ldots, b_s\in\mathcal B$ that allow one to move from $f$ to $g$, visiting only graphs in $\mathcal F_{z,t}$:
\[
g = f +\sum_{i=1}^sb_i \text{ such that }b_i\in \mathcal B, \;\;\;\text{ and }\;\;\; f+\sum_{i=1}^ib_i\in \mathcal F_{z,t} \text{ for all } i\leq s.
\]
If $\mathcal B$ connects $\mathcal F_{z,t}$ for every possible value of the sufficient statistic $t$, then $\mathcal B$ is said to be a \emph{Markov basis} for the valued SBM with block assignment $z$.
\end{defn}

For any log-linear model, the existence and finiteness of a Markov basis are guaranteed by a fundamental result from algebra called the Hilbert basis theorem. It is also well known that while theoretically sound, the Markov bases approach to sampling from conditional distributions may suffer from some computational hurdles in practice. These have been addressed by various authors in the past two decades; see \cite{MB25years} and references therein for a recent overview, and recent work on how to circumvent Markov bases Markov chain convergence issues in general \cite{multilevelAlgStat}. 
In addition, when combined with the parallel method of \cite{BesagCliffordExchangableSample}, MCMC samples provide an exchangeable sample from the relevant fiber of the model. 

The most important consequence is that using Markov bases to sample from the fiber leads to an irreducible Markov chain on $\mathcal F_{z,t}$. 
In Algorithm \ref{alg:GoF-known-block} below, the exact conditional $p$-value of $g$ with fixed block assignment $z$ and $T_{z}(g) = t$, is estimated by one such Markov chain Monte Carlo algorithm, where each execution of Step 4 in the algorithm produces a new graph in the fiber $\mathcal F_{z,t}$ using one Markov basis step. 
Markov bases for the models listed in Section~\ref{section:block-models-valued} are described below. The results generalize Theorem 5.4 in \cite{GoFSBMvariants2023}, which provided Markov bases for the classical SBM.  

\begin{prop}\label{prop:Markov_Basis_poissonSBM} Let $z$ be a block assignment and let $T_z:\Z^{\binom{n}{2}}\to \Z^{\binom{n}{2}}$ be the sufficient statistic for the Poisson SBM. For every $1\leq u <v \leq n$ let $\epsilon_{uv}$ be the vector $g=(g_{uv}:1\leq u<v\leq n)\in\N^{\binom{n}{2}}$ with $g_{uv}=1$ and $0$ everywhere else. Then, $\mathcal B=\{\ep_{uv}-\ep_{u'v'}: z_u=z_{u'}, z_v=z_{v'}\}$ is a Markov basis for the Poisson SBM with block assignment $z$.
\end{prop}

The move $m = \ep_{uv}-\ep_{u'v'}$ in $\mathcal B$ represents an interaction switch: replacing one interaction between nodes $u'$ and $v'$ for an interaction between nodes $u$ and $v$.

\begin{proof}[Proof of \ref{prop:Markov_Basis_poissonSBM}] 
Let $z$ be a fixed block assignment and $f, g\in \mathcal F_{z, t}$ be different valued networks. Assume without loss of generality that $\{u,v\}$ is a dyad with $g_{uv}>f_{uv}$ and $z_u=i, z_v=j$. Since $T_z(f)=T_z(g)$, it follows that
\[
\sum_{u'\in B_i, v'\in B_j} g_{u'v'} = \sum_{u'\in B_i, v'\in B_j} f_{u'v'}.
\]
This implies the existence of a dyad $\{u',v'\}$ such that $g_{u'v'}<f_{u'v'}$. Let $b=\ep_{uv}-\ep_{u'v'}$ and observe that the previous observations imply that $f+b\in \mathbb G$, and $||(f+b)-g||_1 = ||f-g||_1-2$. By an inductive argument, we can find a sequence $b_1, \ldots, b_s\in \mathcal B$ satisfying the conditions in Definition~\ref{defn:Markov_basis}.
\end{proof}

\begin{prop}\label{prop:Markov_Basis_labeledSBM} Let $z$ be a block assignment and let $L\in \Z_+$ represent a number of labels. Let $T_z:\Z^{L\binom{n}{2}}\to \Z^{L\binom{n}{2}}$ be the sufficient statistic for the Labeled SBM. For every $1\leq u <v \leq n, \ell\in [L]$ let $\epsilon_{uv}^{(\ell)}$ be the vector $g= (g^{(\ell')}_{u'v'}:1\leq u'<v'\leq n)\in \mathbb \N^{L\binom{n}{2}}$ such that 
\[
g_{u'v'}^{(\ell')}=
\begin{cases}
1, & \text{if } (u', v') = (u, v)\text{ and } \ell'=\ell\\
0, & \text{otherwise}.
\end{cases}
\]
Then, $\mathcal B=\{\ep_{uv}^{(\ell)}+\ep_{u'v'}^{(\ell')}-\ep_{uv}^{(\ell')}-\ep_{u'v'}^{(\ell)}: \ell, \ell'\in [L], z_u=z_{u'}, z_v=z_{v'}\}$ is a Markov basis for the Labeled SBM with block assignment $z$.
\end{prop}

The move $m = \ep_{uv}^{(\ell)}+\ep_{u'v'}^{(\ell')}-\ep_{uv}^{(\ell')}-\ep_{u'v'}^{(\ell)}$ in $\mathcal B$ represents a pair of interaction-type switches: replacing one $\ell'$-type interaction between $u$ and $v$ for an $\ell$-type interaction, and replacing one $\ell$-type interaction from $u'v'$ for an $\ell'$-type interaction.

\begin{proof}[Proof of \ref{prop:Markov_Basis_labeledSBM}] 
Let $z$ be a fixed block assignment and $f, g\in \mathcal F_{z,t}$ be different valued networks. Assume without losing generality that $g_{uv}^{(\ell)} > f_{uv}^{(\ell)}$ where $z_u=i, z_v=j$ and $\ell\in [L]$. Since $f, g\in \mathbb G$, it follows that $\sum_{\ell'=1}^L g_{uv}^{(\ell')}=N=\sum_{\ell'=1}^{L}f_{uv}^{(\ell')}$, implying the existence of $\ell'\in [L]\minus\{\ell\}$ such that $g_{uv}^{(\ell')}<f_{uv}^{(\ell')}$.  Furthermore, since $T_z(f)=T_z(g)$, we have
\[
\sum_{u'\in B_i, v'\in B_j} g_{u'v'}^{(\ell)} = \sum_{u'\in B_i, v'\in B_j} f_{u'v'}^{(\ell)},
\] 
meaning that there exists $u'\in B_i, v'\in B_j$ with $f_{u'v'}^{(\ell)}> g_{u'v'}^{(\ell)}$. Let $m = \ep_{uv}^{(\ell)}+\ep_{u'v'}^{(\ell')}-\ep_{uv}^{(\ell')}-\ep_{u'v'}^{(\ell)}\in \Z^{L\binom{n}{2}}$ and observe that $f+m\in \N^{L\binom{n}{2}}$, and $\sum_{\ell=1}^L{(f+m)_{uv}^{(\ell)}}=\sum_{\ell=1}^Lf_{uv}^{(\ell)}=N_{uv}$. In other words, $f+m\in \mathbb G$. Furthermore, we have

\[
||(f+m)-g||_1=
\begin{cases}
||f-g||_1 - 4, &\text{if } g^{(\ell')}_{u'v'} > f^{(\ell')}_{u'v'}\\
||f-g||_1 - 2, &\text{otherwise.}  
\end{cases}
\]

By an inductive argument, this shows that the set $\mathcal B$ described in the statement of the proposition is a Markov basis for the valued SBM with block assignment $z$.
\end{proof}

\begin{rmk} The Markov bases described above are known as a \emph{distance reducing} in the algebraic statistics literature, see \cite[Ch.6]{AHT2012}. 
\end{rmk}

Instead of pre-computing the Markov bases before running the test of model fit, we construct one basis element at random following the dynamic move construction of \cite{GoF-dynamicMBs2014}.  \cite{Dobra2012} provides an excellent first demonstration of how MCMC fiber sampling based on dynamically generated Markov bases outperforms Sequential Importance Sampling (SIS) for the types of data we consider here. The output $pval$ of Algorithm \ref{alg:GoF-known-block}, computed in Step 8, is a Monte Carlo estimate of the exact conditional
$p$-value from Equation \eqref{eqn:frequentist-p-value}. 
By now, the approach described here is familiar in the algebraic statistics literature (\cite{AHT2012, DSS09}.  

\vspace{.1in}

\begin{algorithm}[H]
\SetAlgoLined
\SetKwInOut{Input}{input}
\SetKwInOut{Output}{output}

\Input{
$g$, an observed graph on $n$ nodes, \\
$z = (z_1, \ldots, z_n)$, a fixed block assignment, \\
Valued SBM specification \eqref{eqn:exponential_family} with sufficient statistics $T_z(\cdot)$ and base measure $h$,\\
$\text{GoF}_{z}(\cdot)$, a goodness-of-fit statistic, \\
\text{numGraphs}, the number of graphs to sample from the fiber $\mathcal F_{z,t}$ where $t=T_{z}(g)$. 
}
\Output{
$p$-value for the hypothesis test that the valued SBM with block assignment $z$ fits $g$ against a general alternative, and the reference sampling distribution.
}
Set $g_0 = g$, the initial point on the fiber to be the valued network\;
\For{$i = 1$ \KwTo \textnormal{numGraphs}}{
    Randomly construct a move $b$ from a Markov basis $\mathcal B$ for the corresponding model\;
    If $b+g_{i-1}\in F_{T_{z}(g)}$, set $g_i=b+g_{i-1}$ with probability $\min\big\{1, \frac{h(b+g_{i-1})}{h(g_{i-1})}\big\}$,  otherwise $g_i=g_{i-1}$\;
    Compute $\text{GoF}_{z}(g_i)$\;
}
Compute $f_\text{pval} := \#\{{i : \text{GoF}_{z}(g_i) \geq \text{GoF}_{z}(g)}\}$\;
Let $pval = \frac{1}{\text{numGraphs}} \cdot f_\text{pval}$\;
Return $pval$ and the sampling distribution $\{\text{GoF}_{z}(g_i)\}_{i=1}^{\text{numGraphs}}$\;
\caption{Goodness-of-fit test for valued SBM with fixed block assignment}
\label{alg:GoF-known-block}
\end{algorithm}

\subsection{Plug-in conditional test with estimated block assignment}
\label{sec:plug-in z-hat test}

Suppose that the underlying distribution of a valued graph $G$ is of the form $P_\theta(G\mid Z=z)$, $\theta\in\Theta$,  for some fixed $z$. The previous section provides a test for some given $z_0$, which may be different from the true block assignment $z$.
Here, we consider a test for a data-driven model hypothesis using a plug-in estimator $\hat z$ of $z$, the goodness-of-fit test with a data-driven hypothesis is given by: 
\[
    H_0: G\sim P_\theta(G\mid Z=\hat z),
\] 
where $\theta\in\Theta$, and $\hat z$ in $H_0$ is understood as a constant.


\smallskip 

To assess how closely an estimator approximates the true block assignment, we introduce the notion of \emph{agreement} that has been considered in the literature (see \cite[Definition 5]{abbe2018community}). 

\begin{defn}
    The agreement between two block assignments $z, z'\in [k]^n$ is defined as 
    \begin{equation}
        A(z, z')= \max_{\sigma\in S_k}\frac{1}{n}\sum_{u=1}^n\mathds{1}\big(\sigma(z_u) = z'_{u}\big),
    \end{equation}
    where $S_k$ is the set of permutations on $[k]$. Whenever $A(z,z')=1$, there exists a permutation $\sigma\in S_k$ such that $\sigma(z_u)=z'_u$ for every $u\in [n]$. In this case, we write $z'=\sigma\cdot z$. 
\end{defn}

\begin{defn}[cf.\ \cite{abbe2018community}, Definition 4]
    Let $G$ be drawn from a valued SBM as in Definition~\ref{def:valued SBM}, Equation~\eqref{eqn:exponential_family}, with parameter vector $\theta$ and a fixed block assignment $z$.  
    An estimator $\hat z=\hat z(G)$ is called \emph{strongly consistent} if $\mathbb{P}(A(z, \hat z)=1)=1-o(1)$, meaning that $\hat z$ is strongly consistent if $A(z, \hat z)=1$ with high probability as $n$ tends to infinity.
\end{defn}

In the setting where the block assignment $z$ is fixed but unknown, and we have an estimator $\hat z$, we can define the \emph{plug-in} $p$-value 
\[
p(\hat z, g)=\mathbb P(\text{GoF}_{\hat z}(G)\geq \text{GoF}_{\hat z}(g)\;|\; T_{\hat z}(g)),
\]
where $\text{GoF}_{\hat z}$ represents a goodness-of-fit test statistic evaluated based on the estimator $\hat z$.
Although using this plug-in $p$-value to test the data-driven hypothesis does not resolve the issue of using the data twice in a finite-sample setting, the following result ensures the validity of the test for the data-driven hypothesis, asymptotically. (In Section \ref{sec:bayesian test}, we consider a Bayesian version of the test that takes into account the uncertanity in the estimator $\hat z$)

\begin{prop}\label{prop:consistent-estimator} Consider a goodness-of-fit statistic satisfying $\text{GoF}_{\tilde{z}}(g)=\text{GoF}_{\sigma\cdot \tilde{z}}(g)$ for any $\tilde{z}\in [k]^n$ and $\sigma\in S_k$. 
Let $G$ be drawn from a valued SBM as in Definition~\ref{def:valued SBM}, Equation~\eqref{eqn:exponential_family}, with parameter vector $\theta$ and a fixed block assignment $z$. 
Let $\hat z=\hat z(G)$ be a strongly consistent estimator, then $\mathbb{P}(p(z, G)=p(\hat z, G))=1-o(1)$ as $n$ tends to infinity. 
\end{prop}

\begin{proof}
Let $G^{(n)}\sim \text{Valued-SBM}(z, \theta^{(n)})$ for every $n$. Then
\begin{align*}
&\mathbb P(p(z^{(n)}, G^{(n)}) = p  (\hat{z}^{(n)}, G^{(n)})) \\ & \geq P(p(z^{(n)}, G^{(n)}) = p(\hat{z}^{(n)}, G^{(n)}) \mid A(z^{(n)}=\hat z^{(n)})=1)\mathbb P(A(z^{(n)}=\hat z^{(n)})=1) \\
& = \mathbb P(A(z^{(n)}=\hat z^{(n)})=1),
\end{align*}
where the last equality follows from the definition of the (plug-in) p-value, the fact that $T_{z}(g)=T_{z}(g')\iff T_{\sigma\cdot z}(g)=T_{\sigma z}(g')$ for any $\sigma\in S_k$, and the property that $\text{GoF}_{\tilde{z}}(g)=\text{GoF}_{\sigma\cdot \tilde{z}}(g)$ for any $\tilde{z}\in [k]^n$ and $\sigma\in S_k$. Since $\hat z$ is a strongly consistent estimator, it follows that $\lim_{n\to \infty}\mathbb P(p(z^{(n)}, G^{(n)}) = p(\hat{z}^{(n)}, G^{(n)}))=1$.
\end{proof}

\begin{rmk}
Each of the goodness-of-fit statistics considered in Section~\ref{sec:GoF statistic} satisfies the conditions assumed in Proposition~\ref{prop:consistent-estimator}. \end{rmk}


\subsection{
Goodness-of-fit test for unknown block valued SBM: partial Bayes}\label{sec:bayesian test}

Consider now the valued SBM with unknown block assignment $Z$, and assume that the number of blocks $k$ is fixed and known. In this case, 
the goodness-of-fit test takes the following form:

\[ 
    H_0:G\sim   \mathbb P_\theta(G\mid Z=z),
\]
against the general alternative. Note that this is a composite null hypothesis, with unknown (nuisance) parameters $\theta$ and $z$. We remind the reader that $\theta$ depends on the block assignment $z$, but to minimize the additional notational burden, we have suppressed this dependence in the notation.

The plug-in $p$-value uses an estimate $\hat z$, which ignores the uncertainty in estimation of $z$, which calls for a Bayesian approach.  This involves integrating the classical $p$-values (obtained  from the conditional test for each fixed value of $z$) over the posterior distribution of $Z$. More specifically, inspired by the discussion in \cite{Meng94} on two interpretations of posterior predictive $p$-values, we take a two-step approach.  
%
%
Namely, we define a `partial-Bayes $p$-value' as follows: 
\begin{equation}\label{eqn:p-value_bayes}
p_{\text{b}}(g)= \sum_{z\in Z_{n, k}}p(z, g)\mathbb P(Z=z\mid g),
\end{equation}
where $p(z, g)$ is the exact conditional (classical) $p$-value assuming that the block assignment is fixed, and $Z_{n,k}$ is the set of all possible block assignments of $n$ nodes into $k$ blocks. 
This is the mean of the conditional $p$-values defined in Equation~\eqref{eqn:frequentist-p-value}, averaging over the posterior distribution $\mathbb P(Z\mid g)$. 
In other words, as Meng interprets the posterior predictive $p$-value as the posterior mean of a classical $p$-value, we apply this interpretation to the unknown (nuisance) parameter $Z$. 
The dependence on the other unknown parameter, $\theta$, is then naturally removed by conditioning on the sufficient statistic for $\theta$. 

Computing the partial Bayesian $p$-value using the two-step approach implied by Equation~\eqref{eqn:p-value_bayes} allows us to use Algorithm~\ref{alg:GoF-known-block} introduced in the previous subsection as a subroutine. The full procedure for the test is outlined in Algorithm~\ref{alg:alg:GoF-latent-block}. 

\begin{algorithm}[H]
\SetAlgoLined
\SetKwInOut{Input}{input}
\SetKwInOut{Output}{output}

\Input{
$g$, an observed graph on $n$ nodes, \\
Valued SBM model specification \eqref{eqn:exponential_family} with sufficient statistics $T_z(\cdot)$, base measure $h$,\\
$\text{GoF}_{Z}(\cdot)$, choice of goodness-of-fit test statistic, \\
\text{numGraphs}, length of each fiber walk.
}
\Output{
$p$-value for the hypothesis test that the chosen model fits $g$ against a general alternative, and the reference sampling distribution(s).
}
Estimate a distribution, $\pi = \mathbb P(Z \mid g)$, of the block assignment $Z$ given $g$\;
Set $\epsilon := 1/m$, where $m$ is the support size of the estimated distribution $\pi$\;
Construct $\hat{\pi}$ from $\pi$ by thresholding: $\hat{\pi} := \{ \hat{ z} : \pi(\hat{z}) > \epsilon \}$\;
Set $numFibers := |{support}(\hat{\pi})|$, the number of distinct block assignments appearing with significant probability\;
\For{$j = 1$ \KwTo \text{numFibers}}{
    Sample a block assignment $\hat{z}^{(j)}$ from the distribution $\hat{\pi}$\;
    Compute $\text{GoF}_{\hat{z}^{(j)}}(g)$\;
    Compute the $j$th value $pval_j$ and sampling distribution $\{\text{GoF}_{\hat{z}^{(j)}}(g_i)\}_{i=0}^{\text{numGraphs}}$ by Algorithm \ref{alg:GoF-known-block} with the following inputs: $g$, $Z = \hat{z}^{(j)}$, $T_{\hat{z}^{(j)}}(\cdot)$, $h$,  $\text{GoF}_{z^{(j)}}(\cdot)$, \text{numGraphs}\;
}
Return $\sum_j \pi(\hat{z}^{(j)}) \cdot pval_j$ and the corresponding sampling distributions $\{\text{GoF}_{\hat{z}^{(j)}}(g_i)\}_{i=0}^{\text{numGraphs}}$
\caption{Goodness-of-fit test for the valued SBM with {\bf latent} block assignment}
\label{alg:alg:GoF-latent-block}
\end{algorithm}
The reader should note that this approach for computing a partial Bayes $p$-value was also taken by \cite[\S 3]{GoFSBMvariants2023} for Bernoulli blockmodels, while the present version applies to a broader model class of non-Bernoulli SBMs. 

The decision on how to implement step 1 of the algorithm, namely estimating a distribution of the block assignments, is intentionally left to the user. We treat such estimation methods as a black box, in theory. We specify one choice for this step in the simulations section \ref{sec:simulations}.

\section{A goodness-of-fit statistic}\label{sec:GoF statistic}

Both of the algorithms for testing model fit---Algorithm~\ref{alg:GoF-known-block} and Algorithm~\ref{alg:alg:GoF-latent-block}---require the specification of a goodness-of-fit statistic $\text{GoF}_{Z}(\cdot)$. In the most general Bayesian setting, it would be more appropriate to define a `discrepancy variable' (cf. \cite{Meng94}), that is, a parameter-dependent test statistic, depending on both $Z$ and $\theta$. We propose such a test statistic that depends on both $z$ and $\theta$. Given the structure of our tests, it is natural to use the maximum likelihood estimate $\hat \theta_{\mathrm{mle}}$ of $\theta$ for a fixed $z$ in the known $z$ case. In the unknown $z$ case, we use the plug in maximum likelihood estimate $\hat \theta(z)_{\mathrm{mle}}$ corresponding to each $z$ sampled from the posterior distribution $\mathbb P(Z\mid g)$. We provide one such choice of a goodness-of-fit test statistic for each  model variant.

First, some required notation. Given a block assignment $z\in [k]^n$, for every $1\leq i \leq j \leq k$ we define $n_i=|B_i|$ and
\[
n_{ij}=
\begin{cases}
n_i n_j & \text{if } i\neq j,\\
\binom{n_i}{2} & \text{ if } i=j.
\end{cases}
\] 
In other words, $n_{ij}$ is the total number of dyads $\{u,v\}$ with $u\in B_i$ and $v\in B_j$.

\paragraph{Poisson SBM.} Given a block assignment $z$ we use the following block-corrected chi-square statistic
\[
\text{GoF}_z(g) = \sum_{u=1}^n\sum_{i=1}^k \frac{(m_{ui}-n_i\hat{\theta}_{z_ui})^2}{n_i\hat{\theta}_{z_ui}},
\]
where $m_{ui}=\sum_{v\in B_i}g_{uv}$ and for every $1\leq i \leq j \leq k$, $\hat{\theta}_{ij}=\frac{T_{z, ij}(g)}{n_{ij}}$ is the maximum likelihood estimate.

\paragraph{Labeled SBM.} Let us observe that the gradient of the log likelihood ratio given a block assignment $z$, evaluated at $\widetilde\theta_{ij}^{(\ell)}$ is given by 
\begin{align*}
\frac{\partial \log\left(\mathbb P_\theta(G=g\mid Z=z)\right)}{\partial \theta_{ij}^{(\ell)}}(\widetilde\theta_{ij}) 
=-n_{ij}\theta_{ij}^{(\ell)} + T_{z, ij}^{(\ell)}(g)
\end{align*}
for any $1\leq i\leq j\leq k$ and $1\leq \ell< L$. Hence, $\hat \theta_{ij}^{(\ell)}=\frac{T_{z, ij}^{(\ell)}(g)}{n_{ij}}$ is the MLE for $\theta_{ij}$. Since the value of $\hat \theta_{ij}$ is constant on any given fiber, it only needs to be computed once when performing the conditional test. In this case, we consider the following goodness-of-fit statistic: 
\[
\text{GoF}_{z}(g)=\chi^2_{BC}(g, z)=\sum_{u=1}^{n}\sum_{i=1}^k\sum_{\ell=1}^{L-1}\frac{(m_{ui}^{(\ell)}-n_i\hat \theta_{z_ui}^{(\ell)})^2}{n_i\hat \theta_{z_ui}^{(\ell)}}
\] 
where $m_{ui}^{(\ell)}=\sum_{v\in B_i\minus \{u\}}g_{uv}$. Under the Labeled SBM, we have the expected value $\mathbb E[m^{(\ell)}_{ui}] = n_i\theta_{z_ui}$, therefore large values of $\chi^2_{BC}(g, z)$, in which we have replaced $\theta_{z_ui}^{(\ell)}$ with the MLE $\hat \theta_{z_ui}^{(\ell)}$, indicate lack of fit.

\medskip 
The reader familiar with the network literature may notice that the GoF statistics for the valued SBM are analogous to those used for the classical SBM proposed in \cite{GoFSBMvariants2023}. 
Of course, there are other options, for example \cite{Zhang2023-lj}, who propose a different approach. They start with the GoF statistics proposed in \cite{GoFSBMvariants2023}, consider the quantities $m_{ui}$, and then condition on the degree of each node, which makes the collection of $m_{ui}$ for each $u$ a multinomial distribution. Then they derive an asymptotic distribution using this multinomial representation. 
This effectively takes care of using the data twice. We solve this problem by taking the Bayesian approach which, while potentially more computationally intensive, is consistent with the structure of the partial-Bayes two-step test for the latent-block valued SBM.  Another potential workaround is discussed in Section~\ref{sec:discussion}. 

\section{Consistency of the MLE for valued SBM with unknown $z$}
\label{section:consistency of mle}
The goodness-of-fit statistic defined in Section \ref{sec:GoF statistic} depends on both $\theta$ and $z$. In the unknown $z$ case, we use the estimate $\hat \theta(\hat z)_{\mathrm{mle}}$ where $\hat z$ is either a plug-in estimate of $z$, or a sample from the posterior distribution $P(Z|G)$. 
In this section, we show that if $\hat z$ is a \emph{weakly consistent estimator} of the true $z$ (see Definition \ref{def:weakconsistency}, then the plug-in estimator $\hat \theta(\hat z)_{\mathrm{mle}}$ is a consistent estimator of $\theta$.

First we state the result of consistency when $z$ is known. In this case, the valued SBM is an exponential family model, and consistency of $\hat \theta(z)_{\mathrm{mle}}$ follows from the properties of exponential family model. But it is instructive to examine the assumptions that are needed for this to be true, so we state this result as the following lemma:
\begin{lemma}[Consistency of $\hat \theta(z)_{\mathrm{mle}}$ with known $z$]
\label{lem:known_z}
Fix $K$. 
For block pair $(i,j)$ let $n_{ij}$ denote the number of dyads:
\[
n_{ij} = 
\begin{cases}
n_i n_j, & i \neq j,\\
\binom{n_i}{2}, & i=j.
\end{cases}
\]
Assume:
\begin{enumerate}
\item $n_{ij}\to\infty$ for all $(i,j)$.
\item for each $(u,v)$, the dyad distribution $\{G_{uv} : z(u)=i, z(v)=j\}$ belongs to a regular exponential family with canonical parameter $\theta_{ij}$, sufficient statistic $T_{z,ij}(G)$ with finite variance, and identifiable mean--parameter map.
\end{enumerate}
Then the MLE $\hat\theta_{ij}(z)$ computed from $T_{z,ij}(g)$ satisfies
\[
\hat\theta_{ij}(z) \;\xrightarrow{p}\; \theta_{ij}, \quad \text{for all } i,j.
\]
\end{lemma}

\begin{proof}
Fix a block pair $(i,j)$. Conditioned on $z$, the variables 
$T_{z,ij}(G) = \sum_{u \in B_i v \in B_j}T_{z,ij}(G_{uv})$ for $u<v$ with $z(u)=i,\,z(v)=j$ are sum of i.i.d.\ random variables, each with mean 
$\mu_{ij}=\mathbb{E}_{\theta_{ij}}[T_{z,ij}(G)]$ and finite variance.
By the weak law of large numbers,
\[
\frac{1}{n_{ij}}T_{z,ij}(g) \;\xrightarrow{p}\; \mu_{ij}.
\]
In exponential families, the MLE $\hat\theta_{ij}(z)$ is equivalently the solution of the mean equation 
$\mathbb{E}_{\theta}[T_{z,ij}(G)] = n_{ij}^{-1}T_{z,ij}(g)$.
By the regularity assumption, the mapping $\mu \mapsto \theta$ is continuous and one-to-one in a neighborhood of $\mu_{ij}$. 
Hence, by the continuous mapping theorem,
\[
\hat\theta_{ij}(z) \;\xrightarrow{p}\; \theta_{ij}.
\]
See for example \citet{vaart1998asymptotic} or \citet{lehmann2006theory}for general statements on MLE consistency in exponential families.
\end{proof}
\begin{rmk}
    Note that the $\theta_{ij}$'s are variationally independent of each other. Hence, to estimate each of them consistently, the minimum assumption needed is $n_{ij} \to \infty$. $n_{ij}$ is the effective sample size available for each parameter $\theta_{ij}$. This assumption simply states that the amount of information for each parameter $\theta_{ij}$ diverges as $n\to\infty$. Without this assumption, consistency cannot be ensured even with a known $z$. As a concrete counterexample: let $k$ be fixed and suppose block 1 has a constant size, $n_1 = c >0$ for all $n$. Then $n_{11} = {c \choose 2}$, which remains bounded as $n\to\infty$, which means that $\theta_{11}$ cannot be consistently estimated. $n_{ij}$ is the effective sample size for each parameter $\theta_{ij}$ (only when $z$ is known), see also \cite{KrivitskyKolaczyk15} on related question of effective sample size in Exponential random graph models.
\end{rmk}
\begin{rmk}
As noted above, assumption (1) (that $n_{ij}\to\infty$ for all block pairs) is the minimal condition ensuring consistency of $\hat\theta_{ij}(z)$.  
A convenient sufficient condition is to assume i.i.d.\ node labels with block probabilities $\pi_i>0$ and fixed $K$.  
Then $n_i/n \to \pi_i >0$ and hence $n_{ij} \asymp \pi_i\pi_j n^2 \to \infty$ automatically.  
Thus assuming $\pi_i>0$ implies Assumption (1).  
This is the standard formulation used in the SBM literature (see \citet{agresti2002categorical,bishop2007discrete} for analogous results in contingency-table settings).
\end{rmk}

Now we state the result of consistency when $z$ is estimated. Before that we need some definitions.

\begin{defn}[Optimal alignment and misclassification fraction]
\label{def:misclassification}
Let $z \in [K]^n$ be the true community assignment and $\hat z \in [K]^n$ an estimator.
Let $S_K$ denote the set of permutations of $\{1,\dots,K\}$.  
Define the \emph{optimal alignment permutation}
\[
\sigma_n \;\in\; \arg\min_{\sigma \in S_K} \;\frac{1}{n}\sum_{u=1}^n \mathbf{1}\{\hat z_u \neq \sigma(z_u)\}.
\]
The corresponding \emph{misclassification fraction} is
\[
\varepsilon_n(\hat z,z) \;=\; \frac{1}{n}\sum_{u=1}^n \mathbf{1}\{\hat z_u \neq \sigma_n(z_u)\}.
\]
\end{defn}

\begin{defn}[Weak consistency, see {\cite{bickel2009nonparametric,abbe2018community}}]
\label{def:weakconsistency}
Let $z = (z_1,\dots,z_n) \in [K]^n$ be the true block assignments and $\hat z$ an estimator.  
The estimator $\hat z$ is \emph{weakly consistent} if $\varepsilon_n(\hat z,z) \xrightarrow{p} 0$ as $n \to \infty$.
\end{defn}
\begin{thm}[Plug-in consistency of $\hat \theta(\hat z)$ under weak consistency of $\hat z$]
\label{thm:plugin}
Let $\hat z$ be an estimator of $z$ and let $\sigma_n$ be the optimal alignment permutation from Definition~\ref{def:misclassification}. 
Suppose:
\begin{enumerate}
\item (weak consistency) $\varepsilon_n(\hat z,z) \xrightarrow{p} 0$;
\item (per-pair growth) $n_{ij} \asymp n^2$ for all $(i,j)$ (this is implied by the assumption $\pi_i > 0$ for all $i$, see Remark above); 
\item (regularity) for each block pair $(i,j)$ the dyad distribution is a regular exponential family with finite first moment $\mathbb{E}|T(G_{uv})|<\infty$.
\end{enumerate}
Then, for every block pair $(i,j)$,
\[
\hat\theta_{ij}(\hat z) \;\xrightarrow{p}\; \theta_{ij},
\]
where $\hat\theta_{ij}(\hat z)$ is the MLE computed from $T_{\hat z,ij}(g)$.  
All statements are up to the alignment $\sigma_n$.
\end{thm}

\begin{proof}
Write $\dot z := \sigma_n(z)$ for the aligned true labeling. 
Then $\varepsilon_n = n^{-1}\#\{u:\hat z_u\ne \dot z_u\}$ is the fraction of misclassified nodes.
Let $M_n = \{u: \hat z_u\ne \dot z_u\}$ and $m_n=|M_n|=n\varepsilon_n$. Changing from $\dot z$ to $\hat z$ can affect only dyads incident to nodes in $M_n$. 
Let $D_n$ be the number of dyads $\{u,v\}$ that involve at least one misclassified node, then $D_n$ is equal to the total number of dyads minus the dyads where both endpoints  are correctly classified. Thus, $$D_n = {n \choose 2} - {{n-m_n} \choose 2} = nm_n - \frac{m_n^2 + m_n}{2} \leq nm_n = O(n^2 \epsilon_n)$$

The change in $T_{z,ij}(g)$ is then bounded by the sum of $|T(G_{uv})|$ over at most $D_n$ dyads. 
Under Assumption (3), $\mathbb{E}|T(G_{uv})|<\infty$. 
Normalizing by $n_{ij}\asymp n^2$ (Assumption (2)), we obtain
\[
\frac{1}{n_{ij}}\big|T_{\hat z,ij}(g) - T_{\dot z,ij}(g)\big| = O_p(\varepsilon_n).
\]
Since $\varepsilon_n \to 0$ in probability, this normalized difference vanishes in probability. 
By continuity of the exponential-family mean--parameter map, 
\[
\hat\theta_{ij}(\hat z) - \hat\theta_{ij}(\dot z) \xrightarrow{p} 0.
\]
Finally, by Lemma~\ref{lem:known_z}, $\hat\theta_{ij}(\dot z)\xrightarrow{p}\theta_{ij}$. 
Therefore $\hat\theta_{ij}(\hat z)\xrightarrow{p}\theta_{ij}$.
\end{proof}
\begin{rmk}[Comparison with existing literature]
\label{rem:lit_compare}
The implication ``vanishing label error $\varepsilon_n\to0$ $\Rightarrow$ plug-in consistency of $\hat\theta(\hat z)$'' is a standard principle in the SBM literature, but it has been developed in different forms under different modeling and asymptotic regimes. 
For Bernoulli SBMs, \citet{bickel2009nonparametric} show convergence of empirical block edge densities, and \citet{choi2012stochastic} establish parameter consistency when the number of classes grows under appropriate block-size conditions. 
Celisse et al.\ \citep{celisse2012consistency} prove joint consistency of $(\hat z,\hat\theta)$ for maximum-likelihood and variational estimators in the Bernoulli SBM; their proofs contain the same decomposition used here but are directed at the specific estimators studied. 
For valued-edge SBMs (Poisson, Gaussian) \citet{mariadassou2010valued} prove consistency of variational estimators for block parameters. 
Surveying these developments, \citet{abbe2018community} emphasizes that weak label consistency (vanishing misclassification fraction) suffices for consistent parameter estimation by averaging over estimated blocks. 
\end{rmk}

\section{Asymptotic power of the Gof statistic under a merged block assignment}
\label{section: merged blocks asymptotics}

\begin{figure}[t]
	\centering
	\begin{tikzpicture}	[scale=.5, auto]
		\node (UL0) at (0,8) {};
		\node (UR0) at (8,8) {};
		\node (LL0) at (0,0) {};
		\node (LR0) at (8,0) {};
		
		\node (umUR0) at (6,8) {};
		\node (umLR0) at (6,2) {};
		\node (umLL0) at (0,2) {};
		\node (umLab0) at (3,5) { $G_{-(k-1),-k}$};
		
		\node (kUR0) at (7,8) {};
		\node (kLR0) at (7,1) {};
		\node (kLL0) at (0,1) {};
		
		\node (UBk0Lab0) at (6.5, 8.5) {\footnotesize $B_{k-1}$};
		\node (UBk1Lab0) at (7.5, 8.5) {\footnotesize $B_k$};
		\node (LBk0Lab0) at (-0.75, 1.3) {\footnotesize $B_{k-1}$};
		\node (LBk1Lab0) at (-0.5, 0.3) {\footnotesize $B_k$};
		
		\node (UL1) at (12,8) {};
		\node (UR1) at (20,8) {};
		\node (LL1) at (12,0) {};
		\node (LR1) at (20,0) {};
		
		\node (umUR1) at (18,8) {};
		\node (umLR1) at (18,2) {};
		\node (umLL1) at (12,2) {};
		\node (umLab1) at (15,5) { $G_{-(k-1),-k}$};
		
		\node (LBk0Lab0) at (11.25, 1) {\footnotesize $\tB_{k-1}$};
		\node (UBk0Lab0) at (19, 8.5) {\footnotesize $\tB_{k-1}$};
		
		\node (arrow) at (10,4) {\Large$\Rightarrow$};
		
		\draw (UL0.center) -- (UR0.center) -- (LR0.center) -- (LL0.center) -- (UL0.center);
		\draw (UL1.center) -- (UR1.center) -- (LR1.center) -- (LL1.center) -- (UL1.center);
		
		\path[fill=black,opacity=0.2] (UL0.center) -- (umUR0.center) -- (umLR0.center) -- (umLL0.center) -- (UL0.center);
		\draw (UL0.center) -- (umUR0.center) -- (umLR0.center) -- (umLL0.center) -- (UL0.center);
		
		\path[fill=red,opacity=0.2]  (UR0.center) -- (kUR0.center) -- (kLR0.center) -- (kLL0.center) -- (LL0.center) -- (LR0.center) -- (UR0.center);
		\draw  (UR0.center) -- (kUR0.center) -- (kLR0.center) -- (kLL0.center) -- (LL0.center) -- (LR0.center) -- (UR0.center);
		
		\path[fill=blue,opacity=0.2]  (umUR0.center) -- (kUR0.center) -- (kLR0.center) -- (kLL0.center) -- (umLL0.center) -- (umLR0.center) -- (umUR0.center);
		\draw (umUR0.center) -- (kUR0.center) -- (kLR0.center) -- (kLL0.center) -- (umLL0.center) -- (umLR0.center) -- (umUR0.center);

		\path[fill=black,opacity=0.2] (UL1.center) -- (umUR1.center) -- (umLR1.center) -- (umLL1.center) -- (UL1.center);
		\draw (UL1.center) -- (umUR1.center) -- (umLR1.center) -- (umLL1.center) -- (UL1.center);
		
		\path[fill=green,opacity=0.2] (UR1.center) -- (umUR1.center) -- (umLR1.center) -- (umLL1.center) -- (LL1.center) -- (LR1.center) -- (UR1.center);
		
	\end{tikzpicture}
	\vspace{4mm}
	\caption{Since block assignments partition $[n]$, the rows and columns of the adjacency matrix $G$ for a graph may be grouped according to this block assignment. Assuming this layout, the figure shows how the merge operation on blocks $B_{k-1}$ and $B_{k}$, relabels nodes in the adjacency matrix to be in block $\tB_{k-1} = B_{k-1} \cup B_k$.}\label{fig:merge-blocks}
\end{figure}
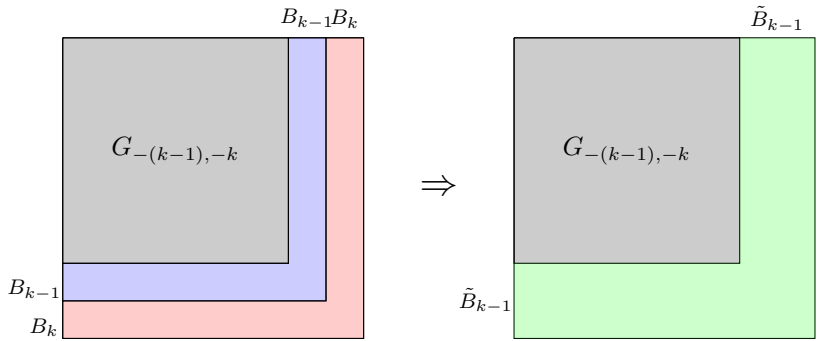

In practice, the number $k$ of blocks  is often not known. In such a case, we show how the goodness-of-fit test proposed here can be used to choose a minimal $k$ for which the test fails to reject the model. We study the asymptotic behavior of the Gof statistic for the labeled SBM in the case of underfitting and show that there is a separation between the underfitting and the true $k$ case, thus elucidating the asymptotic power of the test. The discussion in this section is motivated by the results in \cite{10.1214/16-AOS1457}, where the operation of merging two blocks is performed (see Figure \ref{fig:merge-blocks}) to assess underfitting of a given block assignment and the corresponding $k$. We ignore the case of overfitting, because each $k$-block assignment may be embedded in exponentially many $(k+1)$-block assignments by partitioning a selected block and keeping the dyad parameters the same across the newly created blocks. Furthermore, overfitting the block assignment may lead to situations where $p$-values are artificially large. To see this, consider the extreme example where $k=n$. In this case each block consists of a single node, so the count statistics $T_{i,j}^{(l)}$ are essentially indicators for whether a dyad has the value $l\in [L]$. This means the fiber consists of a single graph, the observed $g$, and the $p$-value is always 1. 

Let $g$ be an observed network on $n$ nodes and $L$ labels, with estimated $k$-block assignment $\hat z$ and dyad parameters $\hat \theta$. Assume that as $n \to \infty$, the estimator $\hat z$ and the true distribution of $g$ are such that the block sizes $|B_1|,\dots,|B_k|$ of $\hat z$ converge in probability; in other words, 
\begin{equation}\label{as:blocksize-dist}
	n_i/n \to \pi_i,\quad \st\quad  \sum_{i=1}^k \pi_i = 1,\text{ and } \pi_i\in (0,1)  \text{ for every } i\in [k].
\end{equation}
This assumption may be justified under the reasoning that if a ground truth $z$ exists for $n$ nodes, then it may be a realization of a random block assignment $Z$, where each $Z(u)$ for $u\in [n]$ follows a categorical distribution with parameters $(\pi_1,\dots,\pi_k)$. A similar assumption, of convergence in probability must be made for $\hat \theta$ as $n\to \infty$:
\begin{equation}\label{as:consistency}
	\hat \theta \to \theta, \quad \st\quad \sum_{l\in [L]}\theta_{i,j}^{(l)} = 1, \text{ and } \theta_{i,j}^{(l)}\in (0,1) \text{ for every } i,j\in [k],\ l\in [L].
\end{equation}
As with the block-size consistency assumption, the convergence of $\hat \theta$ is justified by considering the case when a ground truth $\theta$ exists for some multivariate categorical distribution on the dyad labels on $\hat z$. If this holds, and if we assume $\hat z$ is weakly consistent, then from Theorem \ref{thm:plugin},  the plug-in estimator $\hat \theta(\hat z)$  is consistent for $\theta$.

For an observed $g$, the proposed GoF test compares $g$ to graphs in the fiber $\Fcal_{\hz, t}$ for block assignment $\hz$ with $k$ blocks and sufficient statistics $T_{\hz}(g) = t$. If we consider a new block assignment $\tz$ with $k-1$ blocks (see Remark~\ref{rmk:fiber-subset} below), this affects the sufficient statistic $T_{\tz}(g) = \ttee$ and the corresponding fiber $\Fcal_{\tz, \ttee}$. 
We compare the expected values of the Gof statistic for graphs $g \in \Fcal_{\hz, \hth}$ with graphs $g \in \Fcal_{\tz,\ttee}$. Our goal for this comparison is to give evidence for using a minimal block assignment and the corresponding smallest $k$ over all block assignments for which the GoF test fails to reject the model.

Denote blocks as the sets of nodes $B_i = \{u\in V\colon z_u = i\}$ for $i \in [k]$. A merge operation defines a new block labeling where two blocks $B_i$ and $B_j$ are replaced by their union $B_i \cup B_j$. Without loss of generality, blocks may be relabeled so that $k-1$ and $k$ are the merged blocks. Under this assumption define the merged block assignment to be
\begin{equation}\label{def:merged-block-assignment}
	\tz_u =\begin{cases}
		\hz_u & \text{if } \hz_u < k-1, \\
		k-1 & \text{if } \hz_u \in \{k-1,k\}.
	\end{cases}
\end{equation}
Under the merged block assignment, blocks $\tB_i$ and block sizes $\tn_i$ may be expressed in terms of their counterparts for $\hz$:
\begin{equation}\label{def:merged-blocks}
	\tB_i = \begin{cases}
		B_i, &  i\in [k-2], \\
		B_{k-1}\cup B_{k}, &  i = k-1
	\end{cases}\qquad\tn_i = \begin{cases}
		n_i, & i\in [k-2], \\
		n_{k-1}+n_{k} ,& i = k-1.
	\end{cases}
\end{equation}
The number of dyads can be written similarly:
\begin{equation}\label{def:merged-dyad}
	\tn_{ij} = \begin{cases}
		n_{ij}, &  i,j \in [k-2], \\
		n_{k,j} + n_{k-1,j}, &  i = k-1, j\in [k-2],\\
		n_{i,k} + n_{i,k-1}, &  i \in [k-2], j = k-1,\\
		n_{k,k} + n_{k,k-1} + n_{k-1,k-1}, &  i = j = k-1.
	\end{cases}
\end{equation}
Additionally, the merged label counts for each node $\tm_{u,i}^{(l)}$ and the merged sufficient statistics $T_{\tz,ij}^{(l)}(g) = \ttee_{ij}^{(l)}$ have the form
\begin{equation}\label{def:merged-stats}
	\tm_{u,i}^{(l)} = \begin{cases}
		m_{u,i}^{(l)}, & i \in [k-2] ,\\
		m_{u,k}^{(l)} + m_{u,k-1}^{(l)}, &  i \in i=k-1, \\
	\end{cases}\qquad \ttee_{ij}^{(l)} = \begin{cases}
	t^{(l)}_{ij}, & i,j \in [k-2], \\
	t^{(l)}_{k,j} + t^{(l)}_{k-1,j}, & i = k-1, j\in [k-2],\\
	t^{(l)}_{i,k} + t^{(l)}_{i,k-1}, & i \in [k-2], j = k-1,\\
	t^{(l)}_{k,k} + t^{(l)}_{k,k-1} + t^{(l)}_{k-1,k-1}, &  i = j = k-1.
\end{cases} 
\end{equation}
Finally, the merged parameter estimates $\tth_{ij}^{(l)}$ may be expressed as: 
\begin{equation}\label{def:merged-param}
	\tth_{ij}^{(l)} = \begin{cases}
		\hth^{(l)}_{ij}, & i,j \in [k-2], \\
		\frac{n_k\hth^{(l)}_{k,j} + n_{k-1}\hth^{(l)}_{k-1,j}}{n_k + n_{k-1}}, & i = k-1, j\in [k-2],\\
		\frac{n_k\hth^{(l)}_{i,k} + n_{k-1}\hth^{(l)}_{i,k-1}}{n_k + n_{k-1}}, & i \in [k-2], j = k-1,\\
		\frac{n_{k,k}\hth^{(l)}_{k,k} + n_{k,k-1}\hth^{(l)}_{k,k-1}+ n_{k-1,k-1}\hth^{(l)}_{k-1,k-1}}{n_{k,k} + n_{k,k-1} + n_{k-1,k-1}}, &  i = j = k-1.
	\end{cases} 
\end{equation} 
\begin{rmk}\label{rmk:fiber-subset}
	For $k$-block assignment $(\hz,t)$, and the corresponding merged $(k-1)$-block assignment $(\tz,\ttee)$, $\Fcal_{\hz, t} \subseteq \Fcal_{\tz, \ttee}$ follows directly from \eqref{def:merged-block-assignment} and \eqref{def:merged-stats}. 
    In particular, considering the fiber coming from a merged-blocks block assignment, but taking into account only graphs from the original block assignment, the expectation of GoF statistics will only grow. 
\end{rmk}
For ease of notation we may express some of the above values using the Kronecker $\delta_{ij}$ as follows,
\[n_{ij} = \frac{n_i(n_j- \delta_{ij})}{(1+\delta_{ij})} = \begin{cases}
	n_i n_j, & i \ne j,\\
	\binom{n_i}{2}, & i = j.
\end{cases}\]
Next we consider conditional expectations under fiber membership. 
First we will examine the conditional probabilities of dyad values and then use them to compute the expectation. The following two Lemmas set up the probability and expectation for graphs drawn from the conditional distribution on the fiber. 
In the sequel, we consider a fixed fiber $\Fcal_{\hz,t}$ defined by a fixed $\hat z$ and the corresponding sufficient statistics $t_{\hat z}(g)$. For any graph $g \in \Fcal_{\hz,t}$ $\hat z$ and $t_{\hat z}(g)$ are fixed. Hence, $\hat \theta = f(t)$ is also fixed. Moreover, since $\hat z$ is fixed, $n_i$, the number of nodes in block $i$ and $n_{ij}$, the number of dyads in block $i$ and $j$ are also fixed. Similarly, since $\tilde z$ and $\tilde t$ are fixed, $\tilde \theta = f(\tilde t)$, $\tilde n_i$, the number of nodes in block $i$ under the block assignment $\tilde z$, and $\tilde n_{ij}$ the number of dyads between block $\tilde B_i$ and $\tilde B_j$ are also fixed. In the following conditional expectations, we use this fact and treat $\hat \theta$, $n_i$, $n_{ij}$, $t$, $\tilde z$, $\tilde t$, $\tilde \theta$, $\tilde n_i$ and $\tilde n_{ij}$ as constants, unless stated otherwise.
\begin{lemma}\label{lemma:dyad-cond-prob}
	Suppose $G\sim \PRVSBM(n,\hz, \theta)$. For $l\in [L]$, $i,j\in [k]$ and $u,v\in [n]$ with $\hz_u = i$, $\hz_v = j$. 
    Then $\prob\left(G_{uv}^{(l)}=1 \mid G\in \Fcal_{\hz, t}\right) = \hth_{ij}^{(l)}$, and for $v,v^\prime \in B_i\backslash\{u\}$ where $v\ne v^\prime$:
	\[\prob\left(G_{uv}^{(l)}=1, G_{uv^\prime}^{(l)}=1\,\big|\,G\in \Fcal_{\hz, t}\right)= \frac{n_{ij}\left(\hth_{ij}^{(l)}\right)^2 -\hth_{ij}^{(l)}}{n_{ij}-1}.\]
\end{lemma}
\begin{proof}
	The probability of observing $T_{\hz}(g)=t$ given the $k$-block assignment $\hz$ is  
	\[
	\prob(t\mid \hz) = \prod_{a=1}^k\prod_{b=a}^k\binom{n_{ab}}{t^{(1)}_{ab},\dots, t^{(L)}_{ab}} \prod_{l=1}^{L} \left(\theta_{ab}^{(l)}\right)^{t_{ab}^{(l)}} = p_{-ij}\binom{n_{ij}}{t^{(1)}_{ij},\dots, t^{(L)}_{ij}} \prod_{l=1}^{L} \left(\theta_{ij}^{(l)}\right)^{t_{ij}^{(l)}}, 
	\]
	where $p_{-ij}$ is the marginal probability of the values of $t$ on non $ij$ dyads. We also have joint probability
	\[\prob(G_{uv}^{(l)}=1, t\mid \hz) = \prob(G_{uv}^{(l)}=1\mid \hz)\prob(t\mid G_{uv}^{(l)}=1, \hz) = p_{-ij}\binom{n_{ij}-1}{t^{(1)}_{ij},\dots,(t^{(l)}_{ij} -1),\dots, t^{(L)}_{ij}} \prod_{l=1}^{L} \left(\theta_{ij}^{(l)}\right)^{t_{ij}^{(l)}}.\]
	Then
	\[\prob\left(G_{uv}^{(l)}=1 \mid G\in \Fcal_{\hz, t}\right)=\prob(G_{uv}^{(l)}=1 \mid \hz, t) = \frac{\binom{n_{ij}-1}{t^{(1)}_{ij},\dots,(t^{(l)}_{ij} -1),\dots, t^{(L)}_{ij}}}{\binom{n_{ij}}{t^{(1)}_{ij},\dots, t^{(L)}_{ij}}} = \frac{t_{ij}^{(l)}}{n_{ij}} = \hth_{ij}^{(l)}.\]
	Similarly,
	\[\prob(G_{uv^\prime}^{(l)}=1, t\,\mid G_{uv}^{(l)}=1,  \hz) = p_{-ij}\binom{n_{ij}-2}{t^{(1)}_{ij},\dots,(t^{(l)}_{ij} -2),\dots, t^{(L)}_{ij}}\left(\theta_{ij}^{(l)}\right)^{t_{ij}^{(l)}-1}\prod_{l^\prime \in [L]\backslash\{l\}} \left(\theta_{ij}^{(l^\prime)}\right)^{t_{ij}^{(l^\prime)}}\]
	and 
	\[\prob(t \mid G_{uv}^{(l)}=1, \hz) = p_{-ij}\binom{n_{ij}-1}{t^{(1)}_{ij},\dots,(t^{(l)}_{ij} -1),\dots, t^{(L)}_{ij}}\left(\theta_{ij}^{(l)}\right)^{t_{ij}^{(l)}-1}\prod_{l^\prime \in [L]\backslash\{l\}} \left(\theta_{ij}^{(l^\prime)}\right)^{t_{ij}^{(l^\prime)}}\]
	such that
	\begin{align*}
		\prob\left(G_{uv}^{(l)}=1, G_{uv^\prime}^{(l)}=1\,\big|\,G\in \Fcal_{\hz, t}\right) & = \prob\left(G_{uv}^{(l)}=1\,\big|\,G\in \Fcal_{\hz, t}\right)\prob\left(G_{uv^\prime}^{(l)}=1\,\big|G_{uv}^{(l)}=1,\,G\in \Fcal_{\hz, t}\right) \\
		& = \hth_{ij}^{(l)}\frac{\prob(G_{uv^\prime}^{(l)}=1, t\,\mid G_{uv}^{(l)}=1,  \hz) }{\prob(t \mid G_{uv}^{(l)}=1, \hz)} \\
		& = \hth_{ij}^{(l)} \frac{\binom{n_{ij}-2}{t^{(1)}_{ij},\dots,(t^{(l)}_{ij} -2),\dots, t^{(L)}_{ij}}}{\binom{n_{ij}-1}{t^{(1)}_{ij},\dots,(t^{(l)}_{ij} -1),\dots, t^{(L)}_{ij}}} \\
		& = \hth_{ij}^{(l)} \left(\frac{t_{ij}^{(l)}-1}{n_{ij}-1} \right) = \frac{n_{ij}\left(\hth_{ij}^{(l)}\right)^2 -\hth_{ij}^{(l)}}{n_{ij}-1}.
	\end{align*}
\end{proof}
\begin{lemma}\label{lemma:cond-ex-lab-deg}
	For a graph $G\sim \PRVSBM(n,\hz,\theta)$ the following hold for $u \in B_j$
	\begin{align}
		\Ex\left[m_{ui}^{(l)}\mid G\in \Fcal_{\hz, t}\right] & = (n_i - \delta_{ij})\hth_{ij}^{(l)} \label{eq:cond-ex-degree-stat} \\
		\Ex\left[\left(m_{ui}^{(l)}\right)^2\Big|\, G\in \Fcal_{\hz, t}\right] & = (n_i - \delta_{ij})\hth_{ij}^{(l)}\left[1 + \left(\frac{n_i-\delta_{ij} - 1}{n_{ij}-1}\right)\left(n_{ij}\hth_{ij}^{(l)}-1\right)\right]. \label{eq:cond-ex-sq-degree-stat} 
	\end{align}
\end{lemma}
\begin{proof}
	Lemma \ref{lemma:dyad-cond-prob} gives
	\[\Ex\left[m_{ui}^{(l)}\mid G\in \Fcal_{\hz, t}\right] = \sum_{v\in B_i\backslash\{u\}}\prob\left(g_{uv}^{(l)}=1\mid G\in \Fcal_{\hz, t}\right)=(n_i - \delta_{ij})\hth_{ij}^{(l)}.\]
	Also, using lemma \ref{lemma:dyad-cond-prob} for \eqref{eq:cond-ex-sq-degree-stat} we have:
	\begin{align*}
		\Ex\left[\left(m_{ui}^{(l)}\right)^2\Big|\, G\in \Fcal_{\hz, t}\right] & = \Ex\left[\sum_{v\in B_i\backslash\{u\}}\left(g_{uv}^{(l)}\right)^2 + 2\sum_{\substack{v,v^\prime\in B_i\backslash\{u\}\\v\ne v^\prime}}G_{uv}^{(l)}G_{uv^\prime}^{(l)}\,\Bigg|\, G\in \Fcal_{\hz, t}\right] \\
		& = \Ex\left[m_{ui}^{(l)}\mid G\in \Fcal_{\hz, t}\right] +  2\sum_{\substack{v,v^\prime\in B_i\backslash\{u\}\\v\ne v^\prime}}\prob\left(G_{uv}^{(l)}=1,G_{uv^\prime}^{(l)}=1\,\big|\, G\in \Fcal_{\hz, t}\right) \\
		&=(n_i - \delta_{ij})\hth_{ij}^{(l)} + 2\binom{n_i-\delta_{ij}}{2}\left[\frac{n_{ij}\left(\hth_{ij}^{(l)}\right)^2 -\hth_{ij}^{(l)}}{n_{ij}-1}\right] \\
		& = (n_i - \delta_{ij})\hth_{ij}^{(l)}\left[1 + \left(\frac{n_i-\delta_{ij} - 1}{n_{ij}-1}\right)\left(n_{ij}\hth_{ij}^{(l)}-1\right)\right].
	\end{align*}
\end{proof}

The next result states the stochastic order of the conditional expectation of the GoF statistic for graphs conditional on the fiber, when we use the correct number of blocks $k$. 
\begin{thm}\label{thm:GoF-cond-ex}
	For $G \sim \PRVSBM(n, z, \theta)$ with $k$ blocks, and the block size assumption \eqref{as:blocksize-dist} and consistency assumption \eqref{as:consistency},
	\begin{equation}\label{eq:GoF-cond-ex}
		\Ex\left[\chibc(G,\hz,\hth)\mid G\in \Fcal_{\hz, t}\right] = (L-1)kO_p(n). 
	\end{equation}
\end{thm}
\begin{proof}
Substituting the values from Lemma \eqref{lemma:cond-ex-lab-deg} yields the conditional expectation for each term in the GoF statistic.
\begin{align*}
	&\Ex\left[\frac{\left(m_{ui}^{(l)}-(n_i-\delta_{ij})\hth_{ij}^{(l)}\right)^2}{(n_i-\delta_{ij})\hth_{ij}^{(l)}} \,\Biggl|\, G\in \Fcal_{\hz, t}\right] \\
	& = \frac{\Ex\left[\left(m_{ui}^{(l)}\right)^2\,\Big|\, G\in \Fcal_{\hz, t}\right]}{(n_i-\delta_{ij})\hth_{ij}^{(l)}}- 2\Ex\left[m_{ui}^{(l)}\,\big|\, G\in \Fcal_{\hz, t}\right] + (n_i-\delta_{ij})\hth_{ij}^{(l)} \\
	& = \left[1 + \left(\frac{n_i-\delta_{ij} - 1}{n_{ij}-1}\right)\left(n_{ij}\hth_{ij}^{(l)}-1\right)\right] - (n_i-\delta_{ij})\hth_{ij}^{(l)} \\
	& = 1 - \left(\frac{n_i - \delta_{ij} - 1}{n_{ij}-1}\right) + \left[\left(\frac{n_{ij}}{n_{ij}-1}\right)(n_i - \delta_{ij}-1) - (n_i - \delta_{ij})\right]\hth_{ij}^{(l)}. 
\end{align*}
Then the conditional expectation is
\begin{align*}
	&\Ex\left[\chibc(G,\hz,\hth)\mid G\in \Fcal_{\hz, t}\right]\\
	& = \sum_{l=1}^{L}\sum_{i=1}^k\sum_{j=1}^k\sum_{u \in B_j} \left(1 - \left(\frac{n_i - \delta_{ij} - 1}{n_{ij}-1}\right) + \left[\left(\frac{n_{ij}}{n_{ij}-1}\right)(n_i - \delta_{ij}-1) - (n_i - \delta_{ij})\right]\hth_{ij}^{(l)}\right)\\
	& = \sum_{l=1}^{L}\sum_{i=1}^k\sum_{j=1}^k n_j \left(1 - \left(\frac{n_i - \delta_{ij} - 1}{n_{ij}-1}\right) + \left[\left(\frac{n_{ij}}{n_{ij}-1}\right)(n_i - \delta_{ij}-1) - (n_i - \delta_{ij})\right]\hth_{ij}^{(l)}\right)\\
	& = \sum_{l=1}^{L}\sum_{i=1}^{k}\left[n_i\left(1-\frac{2(n_i-2)}{n_i^2-n_i-2}-\frac{n_i}{n_i+1}\hth_{ii}^{(l)}\right) + \sum_{j\in [k]\backslash\{i\}}n_j\left(\left[1-\frac{n_i-1}{n_i n_j -1}\right]+\left[\frac{n_j(n_i-1)}{n_i n_j -1} - 1\right]n_i\hth_{ij}^{(l)}\right)\right]\\
	& = \sum_{l=1}^{L}\sum_{i=1}^{k}\left[\frac{n_i(n_i-1)}{n_i+1} - \frac{n_i^2}{n_i+1}\hth_{ii}^{(l)} +\sum_{j\in [k]\backslash\{i\}} \left(\frac{n_i n_j}{n_i n_j -1}\right)\left(n_j -1 - (n_j -1)\hth_{ij}^{(l)}\right) \right] \\
	& = \sum_{i=1}^k\left[L\left(\frac{n_i(n_i-1)}{n_i+1} \right) - \frac{n_i^2}{n_i+1}+ (L-1)\sum_{j\in [k]\backslash\{i\}}\left(\frac{n_i n_j (n_j-1)}{n_i n_j -1}\right)\right].
\end{align*}

We now derive the stochastic order of the conditional expectation. Note that the conditional expectation is a function of $\hat z$ and $\hat \theta$, which are now treated as random. Recall the assumption, \ref{as:blocksize-dist}, which states that $n_i/n \xrightarrow{p} \pi_i > 0$. We write $A_n \sim B_n$ to denote asymptotic equivalence in probability: $A_n/B_n \xrightarrow{p} 1$. Let
\[
\begin{aligned}
B_i &:= L\frac{n_i(n_i-1)}{n_i+1} - \frac{n_i^2}{n_i+1} + (L-1)\sum_{j\in [k]\setminus\{i\}} \frac{n_i n_j (n_j-1)}{n_i n_j -1} \\
&\sim L\cdot n_i - n_i + (L-1)\sum_{j\neq i} n_j \\
&= (L-1)n_i + (L-1)\sum_{j\neq i} n_j \\
&= (L-1)\sum_{j=1}^k n_j \\
&= (L-1)n.
\end{aligned}
\]
This implies, $\sum_{i=1}^k B_i \sim \sum_{i=1}^k (L-1)n = k(L-1)n,$ so that in $O_p$ notation: $\sum_{i=1}^k B_i = k(L-1)\,O_p(n)$.

\end{proof}
If the observed graph $g \sim \PRVSBM(z,\theta)$ and assumptions \eqref{as:blocksize-dist} and \eqref{as:consistency} hold, then Theorem \ref{thm:GoF-cond-ex} says that 
the expectation of $\chibc(G,\hz, \hth)$ conditional on $G$ being in the $k$-block fiber $\Fcal_{\hz, t}$ scales as $kLO_p(n)$. Suppose we perform a GoF test and fail to reject the model. Since $\Fcal_{\hz,t} \subseteq \Fcal_{\tz, \ttee}$, it is reasonable to question whether the model under the merged block assignment with smaller number of blocks also fits the observed $g$. The next result shows that for a graph drawn from a fiber with $k$ blocks, if we incorrectly assume that there are $k-1$ blocks, the proposed chi-square statistic scales as $\Omega_p(n^2)$.

\begin{thm}\label{thm:underfit-asym}
	For $g \sim \PRVSBM(z,\theta)$ with $k$ blocks, and assumptions \eqref{as:blocksize-dist} and \eqref{as:consistency},
	\begin{equation}\label{eq:GoF-cond-ex-merged}
		\Ex\left[\chibc(g,\tz,\tth)\mid g\in \Fcal_{\hz, t}\right] = \Omega_p(n^2). 
	\end{equation}
\end{thm}
\begin{proof}
	Consider terms of $\Ex\left[\chibc(g,\tz,\tth)\mid g\in\Fcal_{\hz,t}\right]$ where $i < k-1$ and $u \in B_{k-1}$
	\begin{align}
		& \Ex\left[\frac{\left(\tm_{u,i}^{(l)}- \tn_{i} \tth_{i,k-1}^{(l)}\right)^2}{\tn_{i}\tth_{i,k-1}^{(l)}}\,\Bigg|\,g\in\Fcal_{\hz,t}\right] = \Ex\left[\frac{\left(m_{u,i}^{(l)}- n_{i} \frac{n_k \hth_{i,k}^{(l)} + n_{k-1}\hth_{i,k-1}^{(l)}}{n_k + n_{k-1}}\right)^2}{n_{i} \frac{n_k \hth_{i,k}^{(l)} + n_{k-1}\hth_{i,k-1}^{(l)}}{n_k + n_{k-1}}}\,\Biggl|\,g\in\Fcal_{\hz,t}\right] \nonumber\\
		& = \frac{n_k + n_{k-1}}{n_i\left(n_k\hth_{i,k} + n_{k-1}\hth_{i,k-1}\right)}\Ex\left[\left(m_{u,i}^{(l)}\right)^2\,\Big|\, g\in\Fcal_{\hz,t}\right]  - 2\Ex\left[m_{u,i}^{(l)}\,\Big|\, g\in\Fcal_{\hz,t}\right] + n_{i} \frac{n_k \hth_{i,k}^{(l)} + n_{k-1}\hth_{i,k-1}^{(l)}}{n_k + n_{k-1}} \label{eq:cond-ex-terms}
	\end{align}
	Substituting the conditional expectations from lemma \ref{lemma:cond-ex-lab-deg} into \eqref{eq:cond-ex-terms} gives
	\begin{align}
		&\frac{n_k + n_{k-1}}{n_i\left(n_k\hth_{i,k} + n_{k-1}\hth_{i,k-1}\right)}\left(n_i\hth_{i,k-1}^{(l)}\left[1 + \left(\frac{n_i - 1}{n_i n_{k-1}-1}\right)\left(n_i n_{k-1}\hth_{i,k-1}^{(l)}-1\right)\right]\right) \nonumber\\
		&\quad - 2n_i\hth_{i,k-1}^{(l)} + n_{i} \frac{n_k \hth_{i,k}^{(l)} + n_{k-1}\hth_{i,k-1}^{(l)}}{n_k + n_{k-1}} \nonumber\\
		& = \left(\frac{n_k\hth_{i,k-1}^{(l)} + n_{k-1}\hth_{i,k-1}^{(l)}}{n_k\hth_{i,k} + n_{k-1}\hth_{i,k-1}}\right)\left[1 + \left(\frac{n_i - 1}{n_i n_{k-1}-1}\right)\left(n_i n_{k-1}\hth_{i,k-1}^{(l)}-1\right)\right] \nonumber\\
		& \quad - n_i\left(\frac{n_k\hth_{i,k-1}^{(l)}-n_k\hth_{i,k}^{(l)}}{n_k+n_{k-1}}\right) \label{eq:k-minus-one-terms}.
	\end{align}
	Similarly, for $u^\prime \in B_k$ 
	\begin{align}
		\Ex\left[\frac{\left(\tm_{u^\prime,i}^{(l)}- \tn_{i} \tth_{i,k}^{(l)}\right)^2}{\tn_{i}\tth_{i,k}^{(l)}}\,\Bigg|\,g\in\Fcal_{\hz,t}\right] & = \left(\frac{n_k\hth_{i,k}^{(l)} + n_{k-1}\hth_{i,k}^{(l)}}{n_k\hth_{i,k} + n_{k-1}\hth_{i,k-1}}\right)\left[1 + \left(\frac{n_i - 1}{n_i n_k-1}\right)\left(n_i n_k\hth_{i,k}^{(l)}-1\right)\right] \nonumber\\
		& \quad - n_i\left(\frac{n_{k-1}\hth_{i,k}^{(l)}-n_{k-1}\hth_{i,k-1}^{(l)}}{n_k+n_{k-1}}\right) \label{eq:k-terms}.
	\end{align}
	The terms \eqref{eq:k-minus-one-terms} and \eqref{eq:k-terms} are constant with respect to $u\in B_{k-1}$ and $u^\prime \in B_k$, such that
	\begin{align}
		&\sum_{u\in \tB_{k-1}} \Ex\left[\frac{\left(\tm_{u,i}^{(l)}- \tn_{i} \tth_{i,k-1}^{(l)}\right)^2}{\tn_{i}\tth_{i,k-1}^{(l)}}\,\Bigg|\,g\in\Fcal_{\hz,t}\right] \nonumber\\
		& = \sum_{u\in B_{k-1}} \Ex\left[\frac{\left(\tm_{u,i}^{(l)}- \tn_{i} \tth_{i,k-1}^{(l)}\right)^2}{\tn_{i}\tth_{i,k-1}^{(l)}}\,\Bigg|\,g\in\Fcal_{\hz,t}\right] + \sum_{u^\prime\in B_k} \Ex\left[\frac{\left(\tm_{u^\prime,i}^{(l)}- \tn_{i} \tth_{i,k-1}^{(l)}\right)^2}{\tn_{i}\tth_{i,k-1}^{(l)}}\,\Bigg|\,g\in\Fcal_{\hz,t}\right] \nonumber\\
		& =n_{k-1}\left(\frac{n_k\hth_{i,k-1}^{(l)} + n_{k-1}\hth_{i,k-1}^{(l)}}{n_k\hth_{i,k} + n_{k-1}\hth_{i,k-1}}\right)\left[1 + \left(\frac{n_i - 1}{n_i n_{k-1}-1}\right)\left(n_i n_{k-1}\hth_{i,k-1}^{(l)}-1\right)\right] \label{eq:simple-k-minus-one-terms} \\
		&\quad + n_k\left(\frac{n_k\hth_{i,k}^{(l)} + n_{k-1}\hth_{i,k}^{(l)}}{n_k\hth_{i,k} + n_{k-1}\hth_{i,k-1}}\right)\left[1 + \left(\frac{n_i - 1}{n_i n_k-1}\right)\left(n_i n_k\hth_{i,k}^{(l)}-1\right)\right] \label{eq:simple-k-terms}\\
		&\quad - n_{k-1}n_i\left(\frac{n_k\hth_{i,k-1}^{(l)}-n_k\hth_{i,k}^{(l)}}{n_k+n_{k-1}}\right) - n_kn_i\left(\frac{n_{k-1}\hth_{i,k}^{(l)}-n_{k-1}\hth_{i,k-1}^{(l)}}{n_k+n_{k-1}}\right). \label{eq:cancelled-terms}
	\end{align}

The terms in equation \eqref{eq:cancelled-terms} cancel. As before, we study the stochastic order of the remaining terms  the above expression and let $\hat z$ and $\hat \theta$ be random. Let $E$ denote the expression in equation \eqref{eq:simple-k-minus-one-terms}. Since $n_j = \sum_{u}I(\hat z_u = j)$ are Binomial random variables and $\frac{n_j}{n} \to \pi_i >0$ (by assumption \ref{as:blocksize-dist}), we have $n_j = n \pi_j + o_p(n)$. Similarly, by the consistency of $\hat \theta_{i,j}$ (Assumption \ref{as:consistency}, see also Theorem \ref{thm:plugin}), we have, $\hat \theta_{i,j}^{(l)} = \theta_{i,j}^{(l)} + o_p(1)$. Substituting these in each term of the expression $E$ gives:
\begin{enumerate}
    \item $n_{k-1} = n\pi_{k-1} + o_p(n)$,  
    \item Fraction $\frac{n_k \hat \theta_{i,k-1}^{(l)} + n_{k-1} \hat \theta_{i,k-1}^{(l}}{n_k \hat \theta_{i,k} + n_{k-1} \hat \theta_{i,k-1}} \overset{P}{\rightarrow} \frac{n \pi_k \theta_{i,k-1}^{(l)} + n\pi_{k-1}  \theta_{i,k-1}^{(l}}{n\pi_k \hat \theta_{i,k} + n\pi_{k-1} \hat \theta_{i,k-1}}  = c_1>0$,   
    \item  $1 + \frac{n_i-1}{n_i n_{k-1} - 1}(n_i n_{k-1} \hat \theta_{i,k}^{(l)} -1) \sim 1 + \frac{1}{n_{k-1}}(n_i n_{k-1} \hat \theta_{i,k}^{(l)} -1) \sim n_i  \hat \theta_{i,k}^{(l)} = n \pi_i\theta_{i,k}^{(l)} + o_p(n) $.  
\end{enumerate}
Multiplying these leading-order terms gives $E =  c n^2 + o_p(n^2)$ with $c = c_1 \pi_i \theta_{i,k}^{(l)}>0$.  
 Similarly, one can show that the expression in equation 26 is $cn^2 + o_p(n^2)$. This implies 
 	 \begin{equation*}
 	 	\Ex\left[\chibc(g,\tz,\tth)\mid g\in \Fcal_{\hz, t}\right] \ge \sum_{u\in \tB_{k-1}} \Ex\left[\frac{\left(\tm_{u,i}^{(l)}- \tn_{i} \tth_{i,k-1}^{(l)}\right)^2}{\tn_{i}\tth_{i,k-1}^{(l)}}\,\Bigg|\,g\in\Fcal_{\hz,t}\right] = c n^2 + o_p(n^2) = \Omega_p(n^2).
 	 \end{equation*}
\end{proof}

We can apply Theorem~\ref{thm:underfit-asym} as follows: computing the GoF statistic under the model for $k-1$ blocks for a graph that is actually drawn from a model with $k$ blocks results in a biased statistic. Namely, for large $n$, the GoF statistic satisfies 
$\Ex\left[\chibc(\tG,\tz,\tth)\mid \tG\in \Fcal_{\hz, t}\right]= \Omega_p(n^2)$. This results in a separation between the goodness-of-fit statistic when using the correct number of blocks $k$ versus the incorrect, $k-1$; so that in the first case the value is on the order of $kn$ and in the second case it is on the order of $n^2$. 
In practice, this means that when estimating the number of blocks $k$, we should accept the minimal $k$ for which the GoF test fails to reject the model hypothesis.

\section{Simulations}\label{sec:simulations}

To run Algorithm 2 in practice, one needs to make a choice for the first step: estimating a distribution of the block assignment. In our computations, we estimate the latent block assignments using the variational EM (VEM) algorithm implemented in the \texttt{sbm} R package \citep{sbm_package}, which builds on the variational framework of \citet{mariadassou2010valued}. Specifically, the algorithm maximizes the evidence lower bound (ELBO) under a mean-field approximation of the posterior \(p(z \mid G)\), yielding posterior membership probabilities and corresponding posterior means for each node. While \citet{mariadassou2010valued} do not establish formal consistency guarantees for the estimated labels, such results can be proved along the lines of existing analyses for variational estimators in stochastic block models---see \citet{celisse2012consistency} for the consistency of variational estimators in binary SBMs, \citet{zhao2024dclbm} for mean-field variational inference in Poisson and degree-corrected SBMs and \cite{zhang2020meanfield} for standard SBMs.

\subsection{Simulations for Poisson SBMs - Power and Type 1 error}\label{sec:type I and power}

For the simulations below, we considered four different connectivity matrices $\theta^{(1)}, \theta^{(2)}, \theta^{(3)}, \theta^{(4)}$, shown below. For $r = 2, 4$, all entries $\theta^{(r)}_{ij}$ were drawn independently from the uniform distribution on $[1, 7]$. For $r = 1, 3$, entries $\theta^{(r)}_{ij}$ were drawn uniformly from $[1, 4]$ if $i = j$ and from $[4, 7]$ otherwise.

\[
\begin{aligned}
\theta^{(1)} &= 
\begin{bmatrix}
5.637 & 1.607 & 3.741 & 3.735 \\
1.607 & 5.669 & 2.942 & 1.120 \\
3.741 & 2.942 & 6.084 & 3.232 \\
3.735 & 1.120 & 3.232 & 4.206
\end{bmatrix}, \quad
\theta^{(2)} = 
\begin{bmatrix}
5.173 & 4.938 & 5.790 & 3.162 \\
4.938 & 6.501 & 6.232 & 4.974 \\
5.790 & 6.232 & 3.743 & 2.556 \\
3.162 & 4.974 & 2.556 & 1.364
\end{bmatrix}, \\[1em]
\theta^{(3)} &=
\begin{bmatrix}
4.888 & 2.707 & 1.477 & 1.923 \\
2.707 & 6.385 & 3.898 & 1.102 \\
1.477 & 3.898 & 5.430 & 3.299 \\
1.923 & 1.102 & 3.299 & 5.758
\end{bmatrix}, \quad
\theta^{(4)} =
\begin{bmatrix}
1.274 & 5.378 & 5.292 & 6.570 \\
5.378 & 1.176 & 4.530 & 6.778 \\
5.292 & 4.530 & 2.825 & 4.350 \\
6.570 & 6.778 & 4.350 & 4.015
\end{bmatrix}.
\end{aligned}
\]

For each of $n = 50$ and $n = 100$ nodes, we conducted four power analyses. Fixing $\theta = \theta^{(r)}$ for $r = 1, \dots, 4$, we simulated 100 graphs with $n$ nodes from the Poisson-SBM with block assignment $Z \sim \text{Multinomial}(\bm{\pi}^{(r)})$ and connectivity matrix $\theta^{(r)}$. The vector $\bm{\pi}^{(r)}$ is a probability vector defined by
\[
\bm{\pi}^{(r)} \propto (a, a^2, a^3, a^4),
\]
where $a = 1$ if $r = 1, 2$, and $a = 0.75$ otherwise.

The results of the power tests using Algorithm~\ref{alg:alg:GoF-latent-block} under a significance level of $0.05$ are presented in Tables~\ref{tab:power50} and~\ref{tab:power100}. These show the proportion of null hypothesis rejections number of blocks specified in the test.

\begin{table}[h!]
\centering
\begin{tabular}{|c|c|c|c|c|}
\hline
$\theta$ & 2 blocks & 3 blocks & 4 blocks & 5 blocks \\
\hline
$\theta^{(1)}$ & 1.00 & 0.59 & 0.05 & 0.01 \\
$\theta^{(2)}$ & 1.00 & 0.66 & 0.03 & 0.03 \\
$\theta^{(3)}$ & 0.88 & 1.00 & 0.07 & 0.04 \\
$\theta^{(4)}$ & 1.00 & 0.99 & 0.06 & 0.03 \\
\hline
\end{tabular}
\caption{Null hypothesis rejection ratios for $n = 50$.}
\label{tab:power50}
\end{table}

\vspace{1em}

\begin{table}[h!]
\centering
\begin{tabular}{|c|c|c|c|c|}
\hline
$\theta$ & 2 blocks & 3 blocks & 4 blocks & 5 blocks \\
\hline
$\theta^{(1)}$ & 1.00 & 0.98 & 0.05 & 0.00 \\
$\theta^{(2)}$ & 1.00 & 1.00 & 0.06 & 0.01 \\
$\theta^{(3)}$ & 1.00 & 1.00 & 0.08 & 0.02 \\
$\theta^{(4)}$ & 1.00 & 1.00 & 0.08 & 0.02 \\
\hline
\end{tabular}
\caption{Null hypothesis rejection ratios for $n = 100$.}
\label{tab:power100}
\end{table}

As expected, both tables show that the rejection ratio of the goodness-of-fit test from Algorithm~2 is close to 1 when using 2 or 3 blocks, and close to 0 when using 4 or 5 blocks.

\subsection{Two species networks}\label{sec:selecting k in species networks}

As an example, we analyzed two undirected and valued networks, where nodes represent parasitic fungal species (\( n = 154 \)) and tree species (\( n = 51 \)), respectively. In these cases, edge counts \( g_{uv} \) correspond to the number of shared host species and the number of shared parasitic species, respectively. The data is available in the \verb | R | package \verb | sbm | \cite{sbm_package}, while the data collection details are described in \cite{Vacher2008}.

After sequentially applying our test to assess whether the data fits a Poisson-SBM, we obtained the results presented in Tables~\ref{tab:tree-species} and~\ref{tab:fungal-species}.

\begin{table}[h]
    \centering
    \begin{tabular}{|c|c|c|c|c|c|c|c|c|}
        \hline
        \textbf{Number of Blocks} & 3--7 & 8--9 & 10 & 11 & 12 & 13 & 14 & 15 \\
        \hline
        \textbf{\( p \)-value} & 0 & .01 & .19 & .68 & .93 & .98 & 1 & 1 \\
        \hline
    \end{tabular}
    \caption{Goodness-of-fit results for the tree species network.}
    \label{tab:tree-species}
\end{table}

\begin{table}[!h]
    \centering
    \begin{tabular}{|c|c|c|c|}
        \hline
        \textbf{Number of Blocks} & 3--17 & 18--21 & 22 \\
        \hline
        \textbf{\( p \)-value} & 0 & .01 & .07 \\
        \hline
    \end{tabular}
    \caption{Goodness-of-fit results for the fungal species network.}
    \label{tab:fungal-species}
\end{table}

These results suggest that the tree species network and the fungal species network are better modeled by a Poisson-SBM with \( 10\) and \( 22 \) blocks, respectively. Our results differ from the ICL criterion used on the Poisson-SBM by \cite{mariadassou2010valued}, which selects \(7\) groups of tree species and \(9\) groups of parasitic fungal species.

\section{Discussion}\label{sec:future} \label{sec:discussion}

We have studied the problem of testing goodness-of-fit of valued stochastic blockmodels. To our knowledge, this paper is the first attempt to connect the algebraic statistics methodology behind finite sample tests to non-Bernoulli network models in the blockmodel setting. 
We show concretely how finite-sample conditional tests for network with a fixed block assignment of nodes can be used to construct a partial-Bayesian test for a network with an unknown block assignment of nodes. 
In particular, our construction  of the partial-Bayes $p$-value as a posterior mean of the classical $p$-value takes inspiration from the \cite{Meng94} interpretation of the posterior predictive $p$-value, but with one key difference. Namely, not only does our test statistic depends on both the block assignment $z$ and model parameters $\theta$, but also $\theta$ depends on $z$. This is why our approach is a two-step process to remove dependence on nuisance parameters $z$ and $\theta$. The dependence of the distribution of the test statistic under the null on the  block assignment $z$ is removed by averaging over the posterior distribution of block assignments, while the dependence of the distribution of the test statistic on the other unknown parameters $\theta$ is removed by conditioning on the sufficient statistics. 
Our proposed testing methodology extends an analogous construction from \cite{GoFSBMvariants2023}, which was developed for networks whose dyads are Bernoulli random variables, i.e., data that are summarized as \emph{simple} graphs. 

The conditional distribution of the underlying exponential families given the sufficient statistics can efficiently be sampled using Markov chain Monte Carlo, provided that Markov bases are known. To this end, a cornerstone of the testing approach is a pair of technical results on Markov bases, Propositions~\ref{prop:Markov_Basis_labeledSBM} and~\ref{prop:Markov_Basis_poissonSBM}. These results imply that, for any observed network, it is possible to devise a \emph{scalable} Markov chain sampler of the fiber, which is the reference set supporting the conditional distribution in the exact test. We provide explicit algorithms that incorporate the statistical development of the partial-Bayes $p$-value computation with the algebraic statistics Markov bases construction. 

\medskip
Our other more technical results consider the application of the model goodness-of-fit test for model selection. This problem makes sense in the case when both $z$ and $k$ are  unknown. Namely, for a given set of $n$ nodes, the number of block assignments into $k$ blocks is finite, so there exists some (possibly many)    block assignments for each $k$ that satisfy:  $z = \mbox{argmax }p(z_k, t)$, where     $p(z_k, t)$ is the $p$-value for the proposed GoF statistic, and $z_k$ is a $k$-block assignment. If the observed graph $g$ fits $P(G|Z)$, then there should exist some $z$ with $k$ blocks such that  $p(z, t)$ is larger than some threshold. 
There is an overfitting problem with this setup; namely, max $p(z_n, t) = 1$ with our current definition of $p$-value: $P(GoF(G) \geq GoF(g) | F_{z,t} )$. This is because when $k = n$, the fiber $F_{z,t}$ contains only the observed graph. This is the essence of Section~\ref{section: merged blocks asymptotics}. The main results therein, Theorems~\ref{thm:GoF-cond-ex} and \ref{thm:underfit-asym}, imply that since the $p$-value is expected to decrease if one keeps merging blocks, the following observation directly follows for model selection. If the observed graph fits $P(G|Z)$, there should exist some minimal  number of blocks $k^*$ such that there are corresponding $z^* = \mbox{argmax } p(z_{k^*}, t)$, 
with $p(z_{k^*}, t)$ greater than the threshold.  
These results hold for any estimator $\hat z$ that is weakly consistent for $z$ and a consistent estimator for $\theta$. We use the maximum likelihood estimator $\hat \theta(\hat z)$, which in Section~\ref{section:consistency of mle} we show is consistent whenever $\hat z$ is weakly consistent. 

\medskip 
In light of our model specifications and flexibility of the testing framework, the following quote from \cite{JiJin-comment2} resonates:  \emph{``We are troubled by the implication that we must choose the number of communities, or that there is one right answer"}. This is a notion with which we agree, but do not amplify  in the main text.  Namely, in the text, we mentioned a `true' block assignment.
After deciding on the choice of a model for $P(G|Z)$, the optimal choice $z^*$ (optimal in the sense of achieving the best fit to the observed data) may not equal the $z$ that we can recognize as a ``true" block assignment in reality. It is natural to adopt $z$ suggested by reality, but any goodness-of-fit testing considers the goodness-of-fit in a statistical sense. 
On the other hand, the use of ``true" block assignment is easily warranted in the case where the blocks and membership are actually known, e.g. faculty in departments in a collaboration network, or political party membership in parliamentary data. This may be a strongest case for using our test developed in Section~\ref{sec:plug-in z-hat test}, because it is testing whether dyads depend on known community membership rather than some estimated community.

\medskip 
In practice, there is some choice on how to implement the  block estimation steps in our algorithms, and addressing the computational challenge of Bayesian methods is something that is generally of interest, and not only to our work.  
There is a growing body of work on algorithms for estimating the block assignment for SBMs; see for example \cite{amini2013pseudo,pati2015optimal,geng2016probabilistic,yan2016convex}.  
Results from \cite{pati2015optimal} demonstrate that the method leads to clustering consistency, which in turn guarantees that the test based on the chi-square-like statistic is asymptotically valid. 
When the number of blocks is unknown, one can use the mixture-of-finite-mixtures (MFM) method for SBM from  \cite{geng2016probabilistic}. This is a method that is provably consistent;  see also \cite{NewmanReinert2016EstimatingNumBlocks} for another algorithm for which there is heuristic evidence of consistency. 
Another  Bayesian method for estimating configurations or block membership in the degree-corrected SBM is provided in  \cite{peng2013bayesian}. 
When the graph grows in size, the MCMC used to estimate the block assignment might suffer from slow convergence, and we can apply deterministic estimators instead. 

An open problem related to this is in the algebraic statistics direction:  how does one develop a composite Markov chain, that can sample multiple fibers--namely, those with nontrivial posterior probability--to estimate the Bayesian $p$-value for the latent labeled SBM? In particular, it would be interesting to extend the related work from \cite{SlavkovicZhuPetrovic-tables} on sampling contingency tables given marginals to the space of conditional tables. There is a clear algebraic conjecture on how one might sample multiple fibers using Markov moves and between-fiber moves which preserve marginal, rather than conditional distributions. The challenge is to develop theoretical guarantees in terms of test asymptotics, power, and control convergence adjusted to the correct Bayesian estimates of the block assignment.  One natural place to start would be to consider a block assignment under a multinomial distribution and adjust the weights of the Markov bases Markov chains appropriately.

Another direction of interest for future work is to incorporate labeled SBMs --non-Bernoulli dyads-- with the recent work of  \cite{JJinNetworkGof2025}, who focus on simple graphs -- Bernoulli dyads.

\paragraph{Algebraic statistics of the Labeled SBM.}
We close with a note for our readers who wish to relate our model results to the algebraic statistics literature. A different way to formulate the labeled SBM as a mixture of log-linear models is by setting $G_{uv}\mid Z=z\sim\text{Multinomial}(p^{(1)}_{uv}, \ldots, p^{(L)}_{uv})$, with

\[
\log p_{uv}^{(\ell)} = \alpha_{z_uz_v}^{(\ell)}+\lambda_{uv} \;\;\text{ for every } 1\leq u<v\leq n, \;1\leq \ell <L,
\]
where $\alpha_{z_uz_v}\in \R$ and $\lambda$ is a normalizing constant ensuring that $\sum_{\ell=1}^L p_{uv}^{(\ell)}=1$. Using this parametrization, it can be seen that given a block assignment $z$ and $p=(p_{uv}^{(\ell)}: 1\leq u<v\leq n, \ell\in[L])$ as defined above, $\log p$ belongs to the linear space spanned by the rows of a matrix $A$ with the following block structure
\[
A=
\underbrace{
\begin{pmatrix}
A_{\text{SBM}} & 0 & \cdots & 0 \\ 
0 & A_{\text{SBM}} & \cdots & 0 \\ 
\vdots & \vdots & \ddots & \vdots \\
0 & 0 & \cdots & A_{\text{SBM}} \\
I & I & \cdots & I
\end{pmatrix}}_{L \mbox{ column blocks}}.
\]
Here, $A_{\text{SBM}}$ is the $\binom{k+1}{2}\times \binom{n}{2}$ binary matrix whose rows are labeled by pairs $\{(i,j): 1\leq i\leq j\leq k\}$, columns are labeled by dyads $\{uv: 1\leq u<v\leq n\}$ and whose $(ij, uv)$ entry equals 1 whenever $z_u=i, z_v=j$, $0$ otherwise. We note that the submatrix $A_{\text{SBM}}$ serves as the design matrix for the classic SBM, i.e., the sufficient statistic introduced in \eqref{eq:suff_stat_bernoulli} is the linear transformation associated to the matrix $A_{\text{SBM}}$.

The matrix $A$ is the design or configuration matrix of the labeled SBM. 
In the algebraic statistics literature, the structure of the matrix $A$ is known as the $L$-th \emph{Lawrence lifting} of $A_{\text{SBM}}$; see \cite{LawrenceLifting2003} and \cite[Section 9.8]{AHT2012}. 

Notice that for a given a block assignment $z$ and a sufficient statistic $t\in\N^{L\binom{k+1}{2}}$, 
\begin{align*}
\mathcal F_{z,t} &= \{g\in \mathbb G: T_z(g)=t\}=\{g\in \N^{L\binom{n}{2}}: A g=t\}.
\end{align*}

This means that the Markov basis derived in Proposition~\ref{prop:Markov_Basis_labeledSBM} is a Markov basis for $A$ in the usual sense of \cite{DS98}. Additionally, $A$ can be understood as the incidence matrix of a $(k+1)$-partite graph satisfying the conditions in \cite[Theorem 1.2]{OHquad99}, concluding that the toric ideal $I_A$ is generated by quadratic binomials, or equivalently, $A$ posseses a quadratic Markov basis, just as we proved in Proposition~\ref{prop:Markov_Basis_labeledSBM}.

\section*{Acknowledgements} 
This work began while the authors were in residence at  the program ‘Algebraic Statistics and Our Changing World’ hosted by the Institute for Mathematical and
Statistical Innovation (IMSI).  The authors are particularly grateful to Souvik Dhara, who introduced us to the censored blockmodel and joined initial discussions on that model with the authors. 

\section*{Funding}
SP is partially supported by the Simons Foundation’s Travel Support for Mathematicians Gift 854770. 
SP and MB were supported by DOE/SC award number 1010629.
IMSI is supported by the National Science Foundation (Grant No. DMS-1929348).  

\bibliography{References}
\bibliographystyle{apalike}

\end{document}